\documentclass[preprint,11pt]{elsarticle}
\usepackage{hyperref}
\usepackage{balance}
\usepackage{multicol}
\usepackage{multirow}
\usepackage{amsmath}
\usepackage{float}
\usepackage{lineno}
\usepackage{soul}

\usepackage{makecell}
\usepackage{color}
 \usepackage{booktabs}
\journal{POLAR technical note}

\newcommand{\vect}[1]{\vec{#1}}
\newcommand{\vecm}[1]{\mathbf{#1}}

\graphicspath{{images/}}

\begin{document}
\begin{frontmatter}

\title{In-flight energy calibration of the space-borne Compton polarimeter POLAR }

%\author[psi,ihep]{Hualin Xiao}
\author[psi,ihep]{Hualin Xiao}
%\ead{hualin.xiao@psi.ch}
\author[psi]{Wojtek Hajdas}
%\ead{wojtek.hajdas@psi.ch}
\author[ihep]{Bobing Wu}
\author[isdc]{Nicolas Produit}
\author[ihep]{Jianchao Sun}
\author[dpnc]{Merlin Kole}
\author[ihep]{Tianwei Bao}
\author[dpnc]{Tancredi Bernasconi}
%\author[poland]{Tadeusz Batsch}
\author[dpnc]{Franck Cadoux}
%\author[ihep]{Junying Chai}
\author[ihep]{Yongwei Dong}
\author[psi]{Ken Egli}
\author[isdc]{Neal Gauvin}
\author[psi]{Reinhold Kramert}
\author[ihep]{Siwei Kong}
%\author[ihep,cas]{Hancheng Li}
\author[ihep]{Lu Li}
\author[ihep,cas]{Zhengheng Li}
\author[ihep]{Jiangtao Liu}
\author[ihep]{Xin Liu}
\author[psi]{Radoslaw Marcinkowski}
%\author[isdc]{Silvio Orsi}
\author[poland]{Dominik K. Rybka}
\author[dpnc]{Martin Pohl}
\author[ihep]{Haoli Shi}
\author[ihep]{Liming Song}
\author[ihep]{Shaolin Xiong}
\author[poland]{Jacek Szabelski}
\author[psi]{Patryk Socha}
\author[ihep]{Ruijie Wang}
%\author[ihep,cas]{Yuanhao Wang}
\author[ihep,cas]{Xing Wen}
\author[dpnc]{Xin Wu}
\author[ihep]{Laiyu Zhang}
\author[cas]{Ping Zhang}
\author[ihep]{Shuangnan Zhang}
\author[ihep]{Xiaofeng Zhang}
\author[ihep]{Yongjie Zhang}
%\author[ihep,cas]{Yi Zhao}
\author[poland]{Anna Zwolinska}

\address[psi]{PSI, 5232 Villigen PSI, Switzerland}
\address[ihep]{Key Laboratory of Particle Astrophysics, Institute of High Energy Physics, Beijing 100049, China}
\address[isdc] {ISDC, University of Geneva,1290 Versoix, Switzerland}
\address[dpnc]{DPNC, University of Geneva, quai Ernest-Ansermet 24, 1205 Geneva, Switzerland}
\address[poland]{The Andrzej Soltan Institute for Nuclear Studies, 69 Hoza str., 00-681 Warsaw, Poland}
\address[cas]{University of Chinese Academy of Sciences, Beijing 100049, China}

\begin{abstract}

POLAR is a compact wide-field space-borne detector for precise measurements of the
linear polarisation of hard X-rays emitted by transient sources in the energy range from 50 keV to 500 keV. 
It consists of a 40$\times$40  array of plastic scintillator bars used as a detection material. 
The bars are  grouped in 25 detector modules.
The energy range sensitivity of POLAR is optimized to match with the prompt emission photons from the gamma-ray bursts (GRBs). 
%Measurements of the GRB polarisation in the prompt emission photons would
%provide unique information on event mechanisms as well as on of GRB jets. 
Polarization measurements of the prompt emission would probe source geometries, emission mechanisms and magnetic structures in GRB jets.
The instrument can also detect hard X-rays from solar flares and be used for precise measurement of their polarisation.
POLAR was launched into a low Earth orbit on-board the Chinese space-lab TG-2 on September 15th, 2016.
To achieve high accuracies in polarisation measurements it is essential to assure both before and during the flight a precise energy calibration.  
Such calibrations  are  performed with four low activity $^{22}$Na radioactive sources placed inside the instrument. 
Energy conversion factors are related to Compton edge positions from the collinear annihilation photons from the sources.
This paper presents main principles of the in-flight calibration, describes studies of the method based on Monte Carlo simulations and
its laboratory verification and finally provides some observation results based on the in-flight data analysis.
\end{abstract}

\begin{keyword}
Gamma-ray burst; polarisation; POLAR; In-flight calibration
\end{keyword}
\end{frontmatter}

\section{Introduction}

Gamma-ray bursts (GRBs) are unpredicted and non-repetitive short flashes of gamma-rays appearing in the sky 
at random time and position and typically lasting from few milliseconds to even several hundred seconds. 
During this period they release a huge amount of energy in the order of $10^{48}$ to $10^{55}$ ergs. 
It is comparable to the rest energy of the sun if the energy release is isotropic. 
Thus, they are regarded as the most energetic events in the universe \cite{Gomboc}.
GRBs are produced at cosmological distances being possibly associated with collapses of massive stars or mergers of compact binary systems \cite{zb1, zb2}.
Since their discovery in the 1960s, thousands of GRBs have been detected by various space-borne  instruments  \cite{batse,integral,swift, bepposax}.
In great details they measured their timing, locations and energy spectra. 
Our understanding of GRBs has already enormously improved as a result of these dedicated measurements and studies (see e.g. Refs.  \cite{Gomboc,review1,reviewpaper1} for recent reviews). 
However, many key questions such as emission mechanisms, geometric structure, magnetic properties or 
nature of their jets are not yet answered. 
Direct polarisation measurements would provide much better understanding as well as possible explanations \cite{zb1,zb2}.

To date, tens of polarisation measurements  in the gamma-ray energy range have been performed reporting polarisation levels from 30\% to 80\% 
(see Refs. eg. \cite{gap1, gap2, astrosat}). 
Most of these measurements have limited statistical significance .  Moreover, presented results frequently do not provide a consistent picture of the GRB polarisation \cite{review_2017, apj2017pol}. 
Thus, it is so far difficult to propose firm answers to pending GRB questions.
Both high-quality data and statistically significant polarisation results from new, dedicated instruments 
are still needed to determine the true GRB nature.

POLAR is a compact space-borne GRB detector for polarization measurements of gamma-rays in the energy range from about 50 keV to 500 keV. 
The instrument was developed by an international collaboration of Switzerland, China and Poland. 
It was launched on September 15th, 2016 on-board the Chinese space-lab TG-2 into a low Earth orbit with 
an altitude around 380 km and an inclination of 42.79$^{\circ}$  for an up to three years long observation period.

POLAR detector uses 1600 segmented plastic  scintillator (PS) bars as the gamma-ray detection material.
It has both a large effective detection area of $\sim$ 80 cm$^2$ and a large field view  of 1/3 of the sky. 
Its main goal i.e. measurements of the linear polarisation in GRBs is realized using Compton scattering. 
Distribution of the azimuthal Compton scattering angle extracted from statistically significant number of gamma-rays contains 
information about their mean polarisation level and polarization direction. 
It is determined using positions of two scintillator bars with the maximum energy depositions left by the incoming gamma-ray.
Reliable energy calibration is crucial for precise reconstruction of energy depositions for each detected event.
Before the launch, the whole energy response of POLAR was carefully calibrated in a series of laboratory test campaigns. 
One used radioactive sources, low energy X-ray generators as well as X-ray beams from the synchrotron radiation facilities \cite{silvio, hlieee,hlesrf,zhangxfgain, merlin}. 
It should be noted that the energy response of the detector is susceptible to multiple factors such as thermal drifts, 
high voltage (HV) variations in power supplies, ageing processes in light transmission systems and radiation effects.
It implies that periodic calibrations of the detector energy response in space are necessary. 
For this purpose four weak $^{22}$Na positron sources  were installed inside POLAR at the innermost edges of the four corner modules.
The collinear annihilation photons selected during offline analysis allow for proper calibration of the whole detector. 
This paper presents the principle of the in-flight calibration, describes studies of the method supported by Monte Carlo simulations and their laboratory verification and finally gives several results based on the in-flight data analysis.

\section{POLAR Instrument}
\subsection{POLAR detector}
\begin{figure*}[htb]
 %\onecolumn
\begin{minipage}{60mm}
 \includegraphics[width=0.8\textwidth]{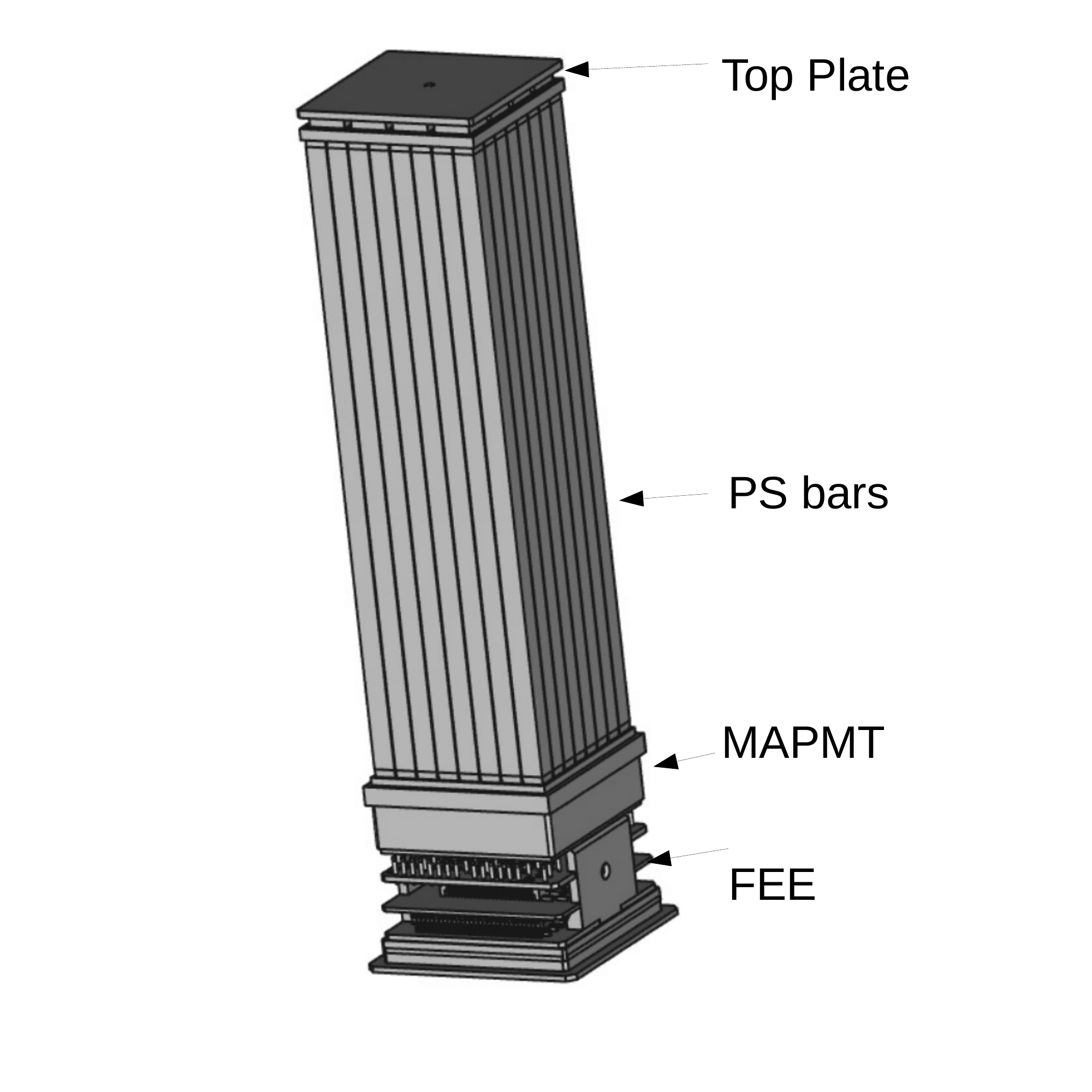}
\end{minipage}
\hspace{\fill}
\begin{minipage}{60mm}
 \includegraphics[width=0.8\textwidth]{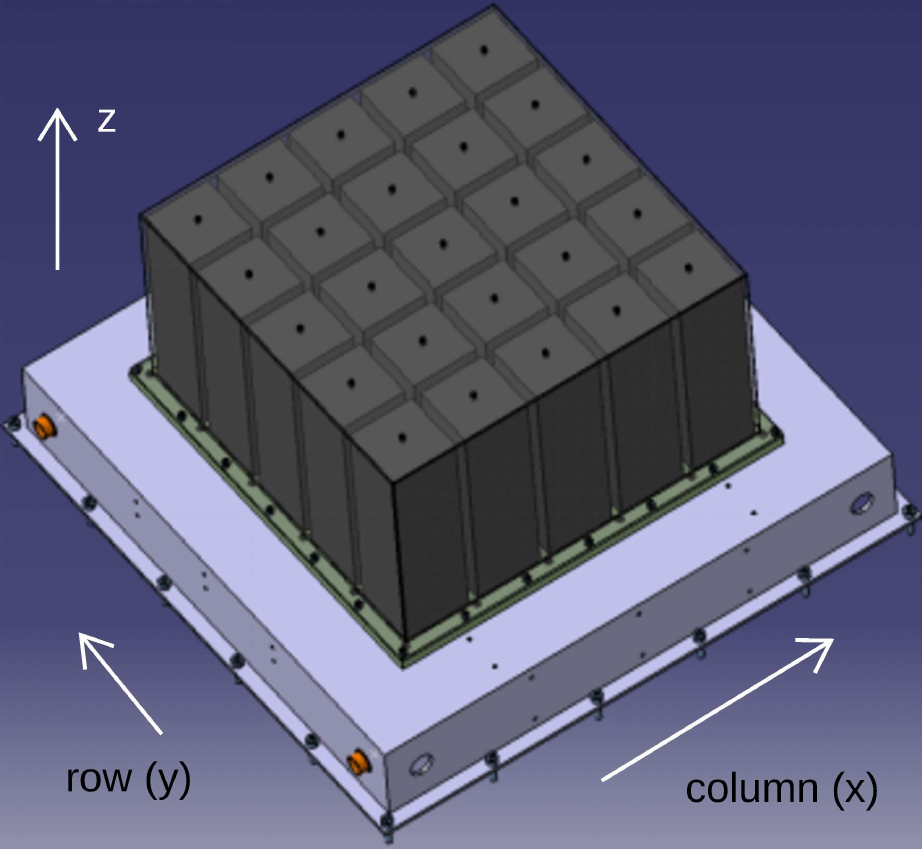}
\end{minipage}
 \caption{ POLAR instrument (OBOX) (right) and structure of its detection module (left).
The full polarimeter consists of 25 such modules. One can also see the coordinate system used in Monte Carlo simulations (right panel).
 }
 \label{fig:polar}
\end{figure*}

POLAR is a hard X-ray Compton polarimeter manufactured using conceptual design
described in Ref. \cite{polarfm}. Polarization measurements are performed using 
angular distribution of the azimuthal angles of the Compton scattered X-rays coming from GRBs. 

As the low-Z materials have a  larger relative Compton scattering cross-sections, 
plastic scintillator (EJ-248M) was chosen as gamma-ray detection target.
It consists of 1600 plastic bars segmented into 25 identical modules as shown in the left panel of Fig.~\ref{fig:polar}.  
Each  module has  8$\times$8 PS bars,  a 64 channel multi-anode photomultiplier MAPMT (Hamamatsu  R10551-00-M64) 
and its own front-end electronics (FEE).
The bar dimensions are 5.8 $\times$ 5.8 $\times$ 176 mm$^{3}$ with both ends cut into a pyramid-like shape in order to match 
the size of the MAPMT pixel and to reduce optical crosstalk between  neighbour bars. 
The surfaces of each bar were polished and wrapped in a highly reflective foil (Vikuiti Enhanced Specular Reflector Film) to increase light collection. 
All bars are coupled to the MAPMT with a 0.7 mm thick optical pad which also partially absorbs vibrations protecting the MAPMT glass.

The FEE consists of three stacked Printed Circuit Boards (PCBs): 
HV divider board, signal processing board and power supply and interfacing board.
The HV divider consists of twelve 470 k$\mathrm{\Omega}$ resistors distributing HV to the MAPMT dynodes. 
The signal processing board contains main electronics with an ASIC chip (IDEAS VA64 with 64 separate readout channels), an ADC, 
a FPGA and a temperature sensor.  
A special internal pulser circuit is also included into it in order to test the gain and the non-linearity of each readout channel. 
The third board has a low voltage power supply circuit and interfacing chips and connectors. 
The whole detector module is packed into a 1 mm thick carbon-fibre socket. 
Specially designed flex cables connect 25 modules to a Central Task Processing Unit (CT) board. 
The CT manages 25 modules and it is also responsible for making trigger decisions, processing data packets, managing the high and the low voltage
power supplies and handling communication with the TG2 space-lab. 
All 25 modules, the CT, the power supplies were placed into an aluminium frame covered with a carbon fibre enclosure (300 $\times$ 300 $\times$ 175 mm$^3$).
The POLAR instrument (OBOX) is shown in the right panel of Fig.~\ref{fig:polar}.
It is mounted on the outside panel of the space-lab facing permanently to the sky. 
A more detailed description the POLAR detector can be found in Ref.~\cite{polarfm}.

\subsection{POLAR OBOX data types}
When ionizing radiation passes through the plastic scintillator, optical photons are emitted as a result of energy deposition.
Generated light pulses are converted by the MAPMT into electric signals that are subsequently shaped and integrated by the ASIC.
Amplitudes of the final electric signal are proportional to the visible energy depositions. 
%If the amplitude of any channel  is above the ASIC discriminator threshold, 
%all 64 analogue signals of the module 
%are held and a trigger signal is sent to the CT starting trigger decision process.
If for any input channel its signal amplitude is above the discriminator threshold of the ASIC, 
a sample-and-hold circuit holds the amplitudes of all 64 channels and a trigger signal is sent to the CT that starts the trigger decision process \cite{vata64}. 
The trigger decision is based on the number of channels triggered within 100 ns. 
If the number is higher than two and smaller than a predefined maximum (usually about 10)
then the event is accepted and the pulse amplitudes subsequently digitized in the ADC \cite{polarfm,polartrig1,polartrig2}. 
Readout data with digitized amplitudes, a timestamp string and some auxiliary information form 
a module science packet transmitted to the CT. 
The event arrival time and the trigger status are also recorded by CT and form a trigger packet. 
Trigger packets are used to identify and merge science packets from different modules belonging to the same event.

Apart from science and trigger packets with the physical data, another two packet types are periodically generated: pedestals and  housekeeping data. 
The CT takes from every module one pedestal event per second. 
Housekeeping data such as information power consumption, operating mode, temperatures of each module and HV values 
is collected by CT and forms a telemetry packet every two seconds.

\section{In-flight calibration requirements}
The main task of POLAR is to provide values of the polarisation degree and polarisation angle for observed GRBs based on
modulation curves constructed for azimuthal Compton scattering angles of detected X-rays.
Energy calibrations are necessary to determine common energy thresholds and identify two bars 
with the maximum energy depositions. 

According to Ref. \cite{hlastro}, the visible energy deposited by gamma-rays in 64 bars of each POLAR module 
can be reconstructed by the following linear transformation: 
\begin{equation}
 \vect{E}_\mathrm{vis}=\vecm{R}^{-1} \vect{E}_{\mathrm{meas}},
 \label{eq:reco}
\end{equation}
where $\vect{E}_\mathrm{vis}$ and $\vect{E}_\mathrm{meas}$ are two vectors representing the
visible and the recorded energy depositions in all 64 bars respectively, and $\vecm{R}$ is  
the response matrix of the module.  $\vecm{R}$ is given  by $ \vecm{R}=\vecm{F}^\mathrm{T} \vecm{M}$, 
where  $\vecm{F}=\left(f_{ij}\right)_{64\times64}$ is the crosstalk matrix and $\vecm{M}=\mathrm{Diag}(m_{0,0},
m_{1,1},...,m_{63,63})$ is the energy conversion matrix whose diagonal element 
is the recorded ADC channel per unit energy deposition in the corresponding bar. 
The transformation allows for energy calibration, including
corrections of crosstalk and non-uniformities. 
Calibration of the crosstalk matrix in Eq.~(\ref{eq:reco}) can be easily done by measuring correlations of signals
between two channels using background data as described in Refs. \cite{silvio, hlieee,hlastro}.  
Thus, the main task of the in-flight calibration is to determine for each module its energy conversion matrix that
consists of 64 energy conversion factors (in units of ADC channel / keV). 
A proper energy calibration in-flight is for several reasons rather complex: 
1) 1600 channels have to be calibrated simultaneously; 
2) plastic scintillators have rather poor energy resolution;
3) background rates are high due to low threshold values and POLAR large field of view; 
4) calibrations should not jeopardise polarisation measurements (implying low source activity). 
Initial conceptual and optimization studies began with a series of Monte Carlo simulations. They were 
conducted to verify several in-flight calibration options such as solar flares, the Crab Nebula and GRBs themselves.
Unfortunately, none of them had clear spectral features useful for energy calibration.
Therefore further studies were performed with various radioactive sources optimised for calibrations in-flight. 
\section{In-flight calibration method}
\subsection{$^{22}$Na calibration sources}
\begin{figure}[H]
\begin{center}
 \includegraphics[width=0.7\textwidth]{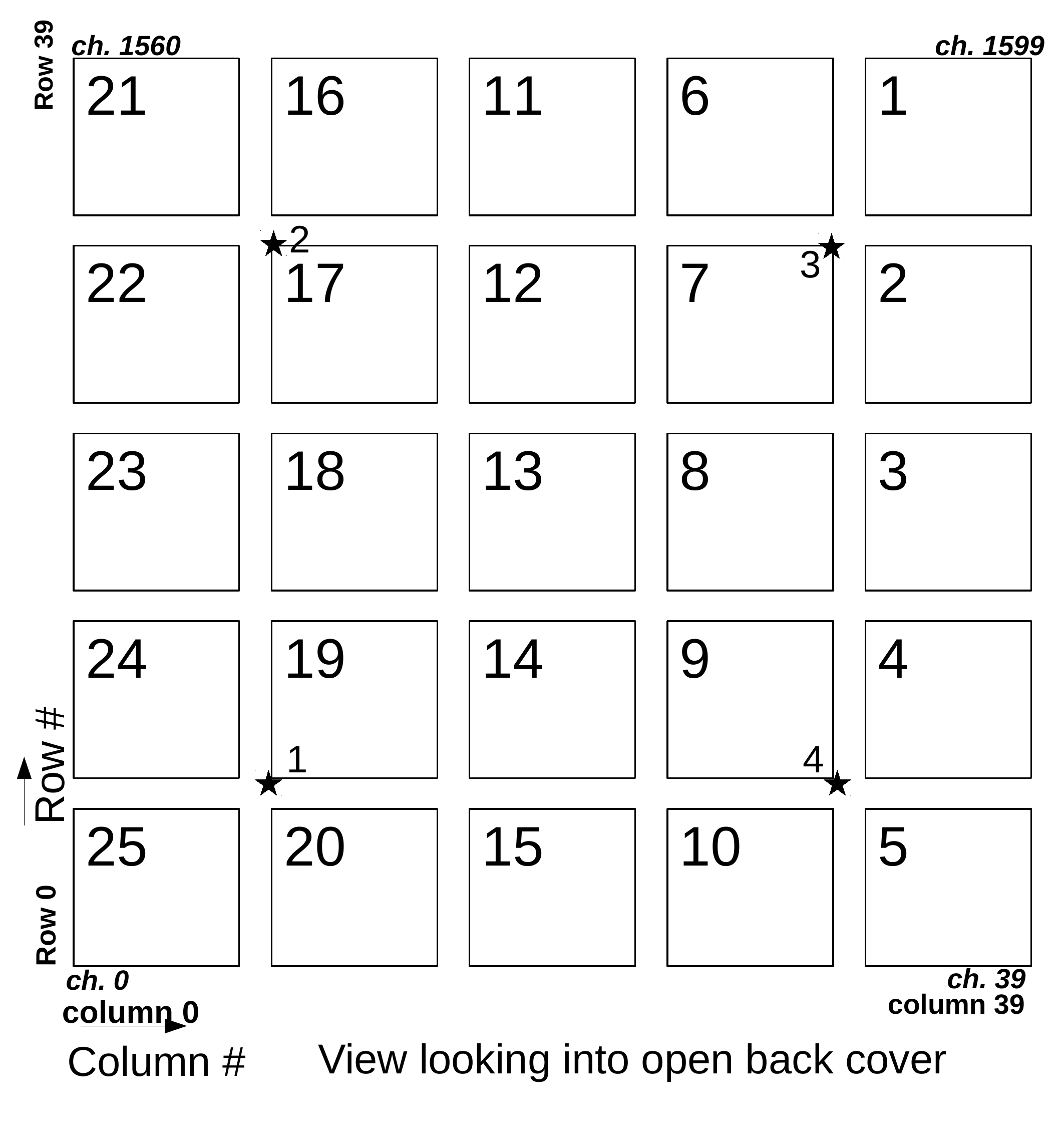}
\caption{Positions of four $^{22}$Na radioactive sources at the corner modules are indicated by stars. 
Source positions, numbering as well as naming of modules and bars
(POLAR channels) and coordinate system conventions are defined and adopted for this paper data analysis. 
}
\label{fig:na22loc}
\end{center}
\end{figure}

A set of four weak $^{22}$Na $\beta^+$ radioactive sources (each with a specific activity value between 100 Bq and 250 Bq) 
was proposed for the in-flight calibration of POLAR. Their configuration was optimized for using collinear 511 keV gamma-rays emitted 
during positron-electron annihilation.
Using four calibration sources allows for higher and more uniform coincidence rates among POLAR scintillator bars. 
The $^{22}$Na isotope was chosen due to its relatively long half-life of 2.6 years as compared with other positron sources \cite{na22data}.
The four  sources are located at the corners of modules 7, 9, 17 and 19 as  shown in Fig.~\ref{fig:na22loc}. They are
 fixed outside the carbon fibre sockets at a half of the bar length about 88 mm from the bar's top surface.
%As shown in Fig.~\ref{fig:na22decay}, 
The $^{22}$Na nucleus decay channel with a positron emission occurs 
a probability of 90.3\% (see Ref. \cite{na22data} for the decay scheme and Fig.~\ref{fig:na22positronspectrum} for the positron emission spectrum).  
The subsequent positron annihilation creates two 511 keV photos flying in opposite directions.
Both photons traverse through the detector having a similar probability of interacting with the scintillator bars located on 
opposite sides of the source. The most probable interaction process is the Compton scattering. 
Corresponding detector hits are attributed to the same event if their time interval is within a hardware 
related coincidence time window of 100 ns.  
The annihilation photons are flying along a line in opposite direction the bars that are hit by them can be accurately 
selected. 
It is applied during offline analysis using conditions based on the event collinearity with two bars and the source.
The coincidence selection accepts $^{22}$Na decay events and could  effectively reject background events from space 
(e.g. diffuse cosmic x-rays, electrons and protons). 
Compton edges in energy spectra created with above conditions provide energy conversion factors for each POLAR channel.

\subsection{Coincidence hit selection algorithm}
\label{sec:selalg}
\begin{figure}[H]
\begin{center}
 \includegraphics[width=0.8\textwidth]{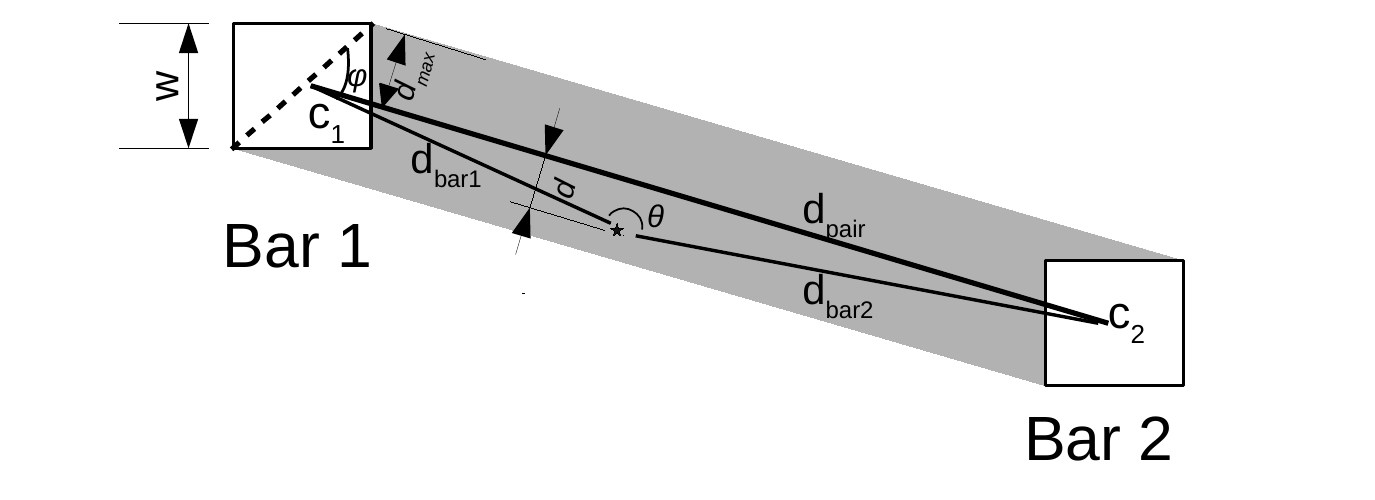}
\caption{Visualization of parameters used to constrain calibration events: perpendicular distance $d$, opening angle $\theta$ and 
angle $\varphi$ between the diagonal line of the bar and the line connecting two bar centres. They define the maximum allowed 
perpendicular distance $d_{\rm max}$. C$_1$ and C$_2$ are the centres of two bars and the $^{22}$Na source is indicated by a star. 
}
\label{fig:coincidence}
\end{center}
\end{figure}  

It is impossible to determine exact locations of of photon interactions inside of the bars.
As shown in Fig.~\ref{fig:coincidence}, two bars can only be hit simultaneously by the collinear annihilation photons  
when the position of the source is inside the polygon depicted as a grey region. 
The perpendicular distance $d$, i.e. the distance from the source to the line connecting two bar centres,
together with the opening angle $\theta$ must satisfy  the conditions  $d<d_{\rm max}$ and $\theta >90$ degree. $d_{\rm max}$ is the maximum distance of the four vertices of one of the bars to the line connecting the two bar centres.
Obviously,  $d_{\rm max}={\rm max}\left(\sqrt{2}w |\sin(\varphi)|/2, \sqrt{2} w |\cos(\varphi) |/2 \right)$, and $ w/2 \leq d_{\rm max} \leq \sqrt{2} w/2$, where $w=5.8$ mm is the width of the PS bar and $\varphi$ is the angle between the diagonal line of the bar section on the XY plane and the line connecting centers of two bars.
Calibration events can be selected by imposing conditions as above to all coincidence hits. 
In practice, the value of $d_{\rm }$ is slightly increased to take into account the size of the source.
An energy spectrum from the hits selected for each bar %can be obtained with a sufficient amount of data. 
shows a clear Compton edge signal observed at the energy of 340.7 keV. 
The identification of Compton edge position in the energy spectrum leads to the energy calibration factor.  
%The coincidence selection condition effectively rejects most of the hits produced e.g. either by the 1.274 
%MeV gamma-rays from the same $^{22}$Na sources or background events from space (e.g. diffuse cosmic x-rays, electrons and protons). 

\section{Monte Carlo simulations}
\subsection{The simulation package}
A complete Monte Carlo simulation package was built based on the GEANT4 suite \cite{geant4} developed by CERN.
Physical processes are described by the `emlivermore\_polar' physics model, which has a high accuracy of
electron, hadron and ion tracking.  
Incident particle definitions use either the General Particle Source toolkit (GPS) from GEANT4 
or particle generators implemented in the POLAR simulation package (e.g. collinear photons and $^{22}$Na events).
Several parameters are recorded for each simulated event including the incident particle type, its energy, 
position, momentum and direction as well as visible energy depositions and positions of each interaction in the scintillator bars. 
Simulation outputs are saved using the ROOT file format\cite{root}.
Extensive set of Monte Carlo simulations has been started already during the R\&D phase of the in-flight calibration method. 

\subsection{Simulations of $^{22}$Na calibration sources} 
\begin{figure}[H]
\begin{center}
 \includegraphics[width=0.7\textwidth]{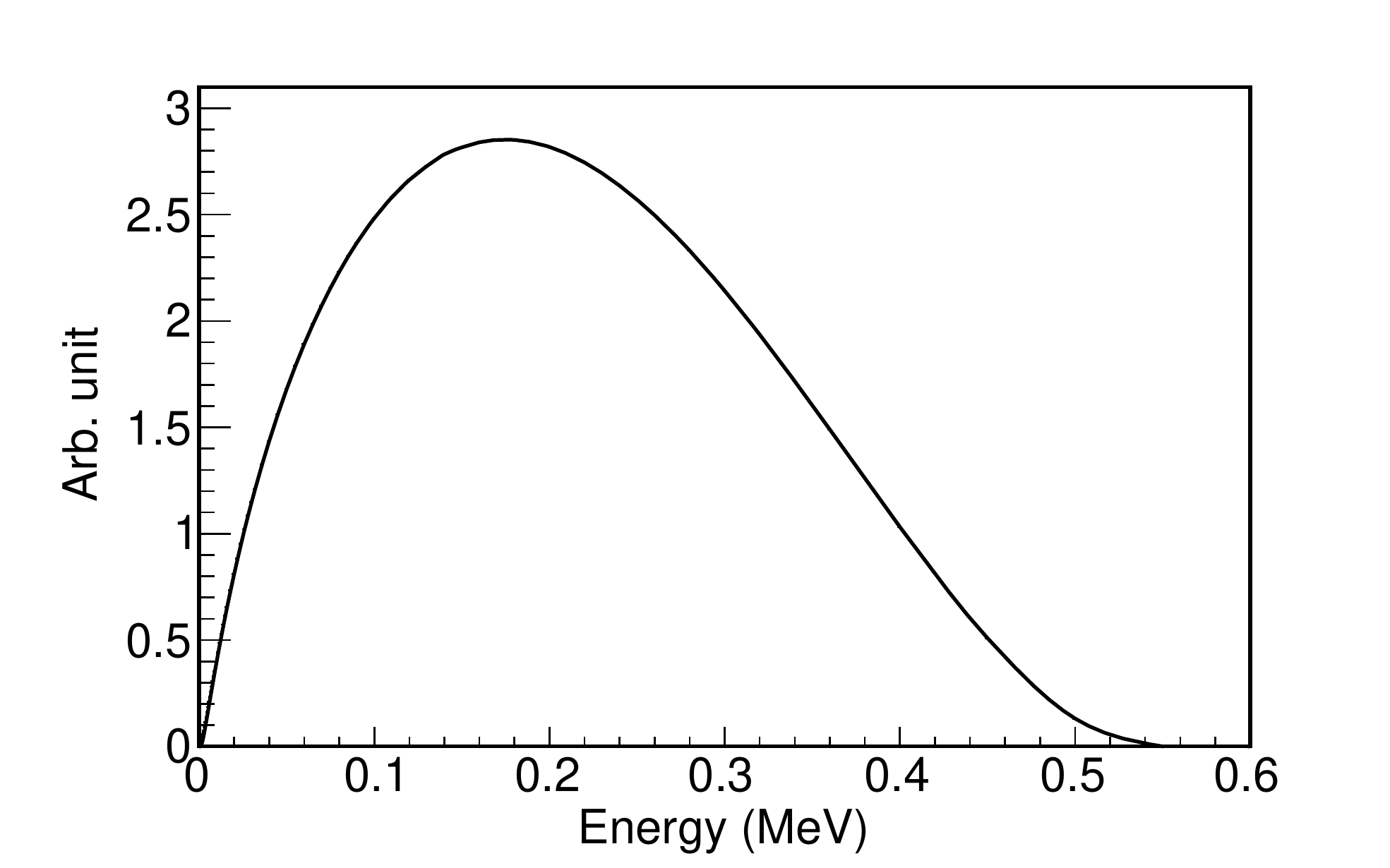}
\caption{ Kinetic energy spectrum of positrons emitted by $^{22}$Na (see Ref.~\cite{icrp38}).
}
\label{fig:na22positronspectrum}
\end{center}
\end{figure}

\begin{figure}[H]
\begin{center}
 \includegraphics[width=0.7\textwidth]{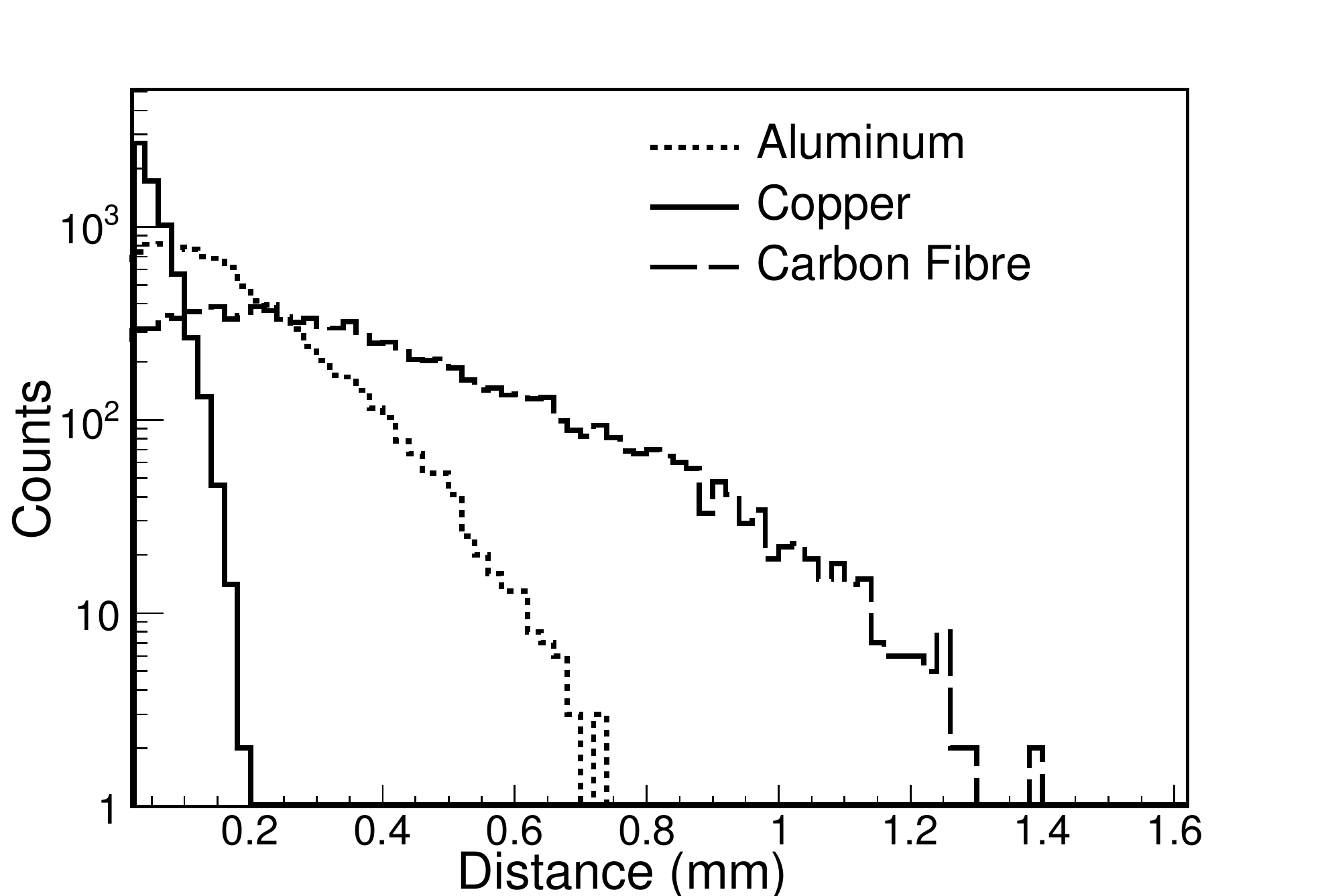}
\caption{ Distributions of $^{22}$Na positron traversing distances
in large blocks of copper, aluminium and carbon fibre. 
The mean transverse distances are 0.04 mm, 0.15 mm, and 0.3 mm respectively. }
\label{fig:positron_free_path}
\end{center}
\end{figure}

The $^{22}$Na source emits gamma-rays with energy of 1274 keV isotropically and with almost 100\% probability.  
It also emits positrons  with  probability of 90.3\% \cite{na22data}.  
The positron energy spectrum is shown in Fig.~\ref{fig:na22positronspectrum}. 
Positrons travel some distances before annihilating with encountered electrons.
Their stopping range depends on the initial energy and the base material of the source structure. 
The minimum thickness needed to stop positrons was estimated for various materials with the help of Monte Carlo simulations. 
As an example, Fig.~\ref{fig:positron_free_path} shows simulated distributions of positron traversing distances 
(i.e. the distances between their initial position and annihilation places) in copper, aluminium and carbon fibre.

The positron initial kinetic energy was sampled from the beta emission spectrum shown in Fig.~\ref{fig:na22positronspectrum}.
The mean distances for the three materials listed above are equal to 40 $\mu$m, 150 $\mu$m and 300 $\mu$m respectively.  
For example, a few hundred $\mu$m thick copper plate can stop almost all positrons. 
Moreover, all values of the stopping lengths are much smaller than the lower limit of $d_{\rm max}$, which is $w/2=2.9$ mm. 
Calibration sources used in flight were glued between two  copper foils having a thickness of 0.5 mm each.
Therefore, for the coincidence hit selection the travelling distances of positrons in the source can be ignored. 
In the following simulations the $^{22}$Na sources are simplified to be point-like.

\subsection{Gamma-ray detection efficiencies}
\label{sec:detection_eff}

\begin{figure}[H]
\begin{center}
 \includegraphics[width=0.7\textwidth]{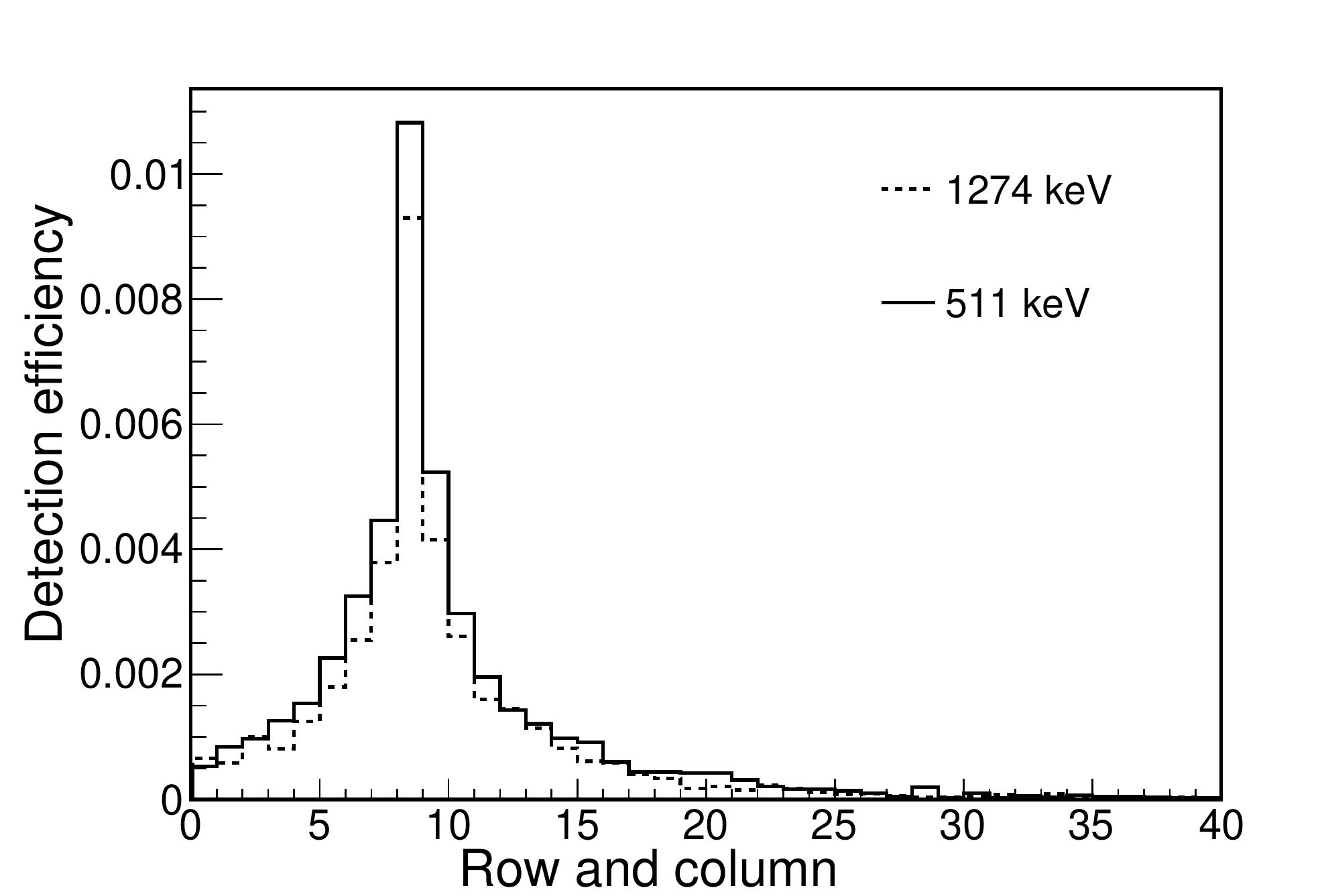}
\caption{Detection efficiencies for gamma-rays of 511 keV and 1274 keV in 40 diagonal bars 
aligned with the source No. 1. Gamma-rays were emitted isotropically and detection threshold values were equal to 5 keV.}
\label{fig:hitsdistr}
\end{center}
\end{figure}

Five simulation runs (see Table~\ref{tb:runs} for details) were performed 
to study efficiency of POLAR for detection of gamma-rays of different energies. 
For the first two runs, gamma-rays with energies of 511 keV and 1274 keV were generated 
at the source 1 position. 
For the third run two annihilation photons flying opposite directions were generated at the same source position.
%#The isotropic angular distribution
The fourth run was used to simulate the $^{22}$Na source 1. For this run the gamma-rays with energy of 1274 keV were generated 
with 100\% probability while two back-to-back annihilation gamma-rays with 90.3\% probability. 
Note that their initial directions were isotropically distributed over 4 $\pi$ sr. 
The purpose of the last run was to estimate the accidental coincidence efficiency.
For this run the gamma-rays with energy of 511 keV were  generated  towards scintillator bars (-z) 
from positions randomly generated on the top surface of OBOX.  
One million events were simulated for each run.

\begin{figure}[H]
\begin{center}
 \includegraphics[width=0.7\textwidth]{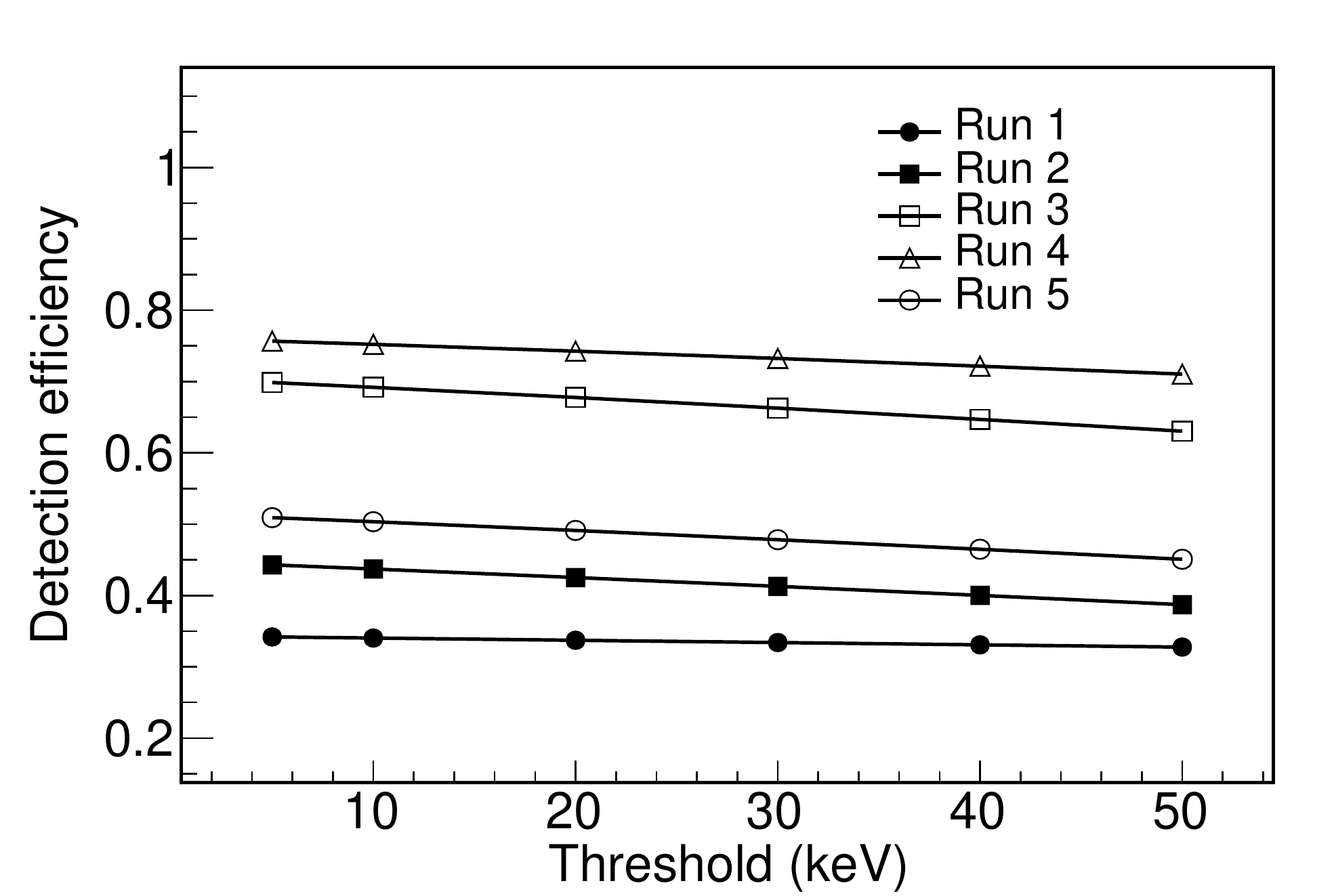}
\caption
{Simulated detection efficiencies of the full POLAR instrument for different gamma-ray sources as a function of deposited energy threshold. }
\label{fig:polar_det_eff}
\end{center}
\end{figure}

Fig.~\ref{fig:hitsdistr} shows detection efficiencies of 40 diagonal bars computed for gamma-rays with energies of 511 keV and 1274 keV
as in the first two runs.  A hit event was counted if its energy deposition in any bar was larger than 5 keV.
Fig.~\ref{fig:polar_det_eff} shows detection efficiencies of the full detector for five different threshold values ranging
from 5 keV to 40 keV.  
An event is considered as detected  if the deposited energy is above
the threshold for at least one bar. 
It can be seen that the detection efficiencies do not significantly depend on the choice of the threshold.
According to the simulations, 34.2\% gamma-rays of 1274 keV and about 44.3\% of 511 keV are detected  at the 5 keV threshold.
Detection probabilities of collinear photon events and a $^{22}$Na event are $\sim$ 70\% and $\sim$ 76\%, respectively.
It is worthwhile to  mention that the efficiencies defined here are  higher than the overall efficiency of POLAR in the context of polarization measurements.
Only events with two or more triggered bars are required to compute an azimuthal scattering angle.

\begin{table}[h]
\centering
\caption{Simulation runs performed to study gamma-ray detection efficiency and coincidence probability. 
Simulated percentages of detected events $f_{\rm d}$ and events with coincidence hit pairs $f_{\rm c}$ are shown. 
The hit selection threshold value was equal to 5 keV.
}
\label{tb:runs}
%\setlength{\tabcolsep}{2pt}
%/opt/geant4/g4work/g4POLAR_201805_forPaper/Na22FinalSim_201804/*.xls
%on polar server
\resizebox{\textwidth}{!}{%
\begin{tabular}{ccccccc}
\toprule
Run \# &  Particle &Energy (keV) & Position & Direction & $f_{\rm d} (\%)$& $f_{\rm c} (\%) $   \\ %dmax=0
\hline
1  & gamma & 1274 & Source 1 & isotropic& 34.2&0.6\\ 
2  & gamma & 511 & Source 1 &  isotropic& 44.3&1.0\\%
3  & collinear  gamma-rays& 2$\times$511 & Source 1 & isotropic& 69.9&17.8\\
4  & $^{22}$Na  & 1274 and 2$\times$ 511 (90.3\% prob.) & Source 1 & isotropic& 75.7&18.2\\
5 &  gamma& 511 & top surface & -z & 50.9& 0.3 \\
\bottomrule
\end{tabular}
}
\end{table}

\subsection{Coincidence selection}
\label{sec:selection}

\begin{figure}[H]
\begin{center}
 \includegraphics[width=0.7\textwidth]{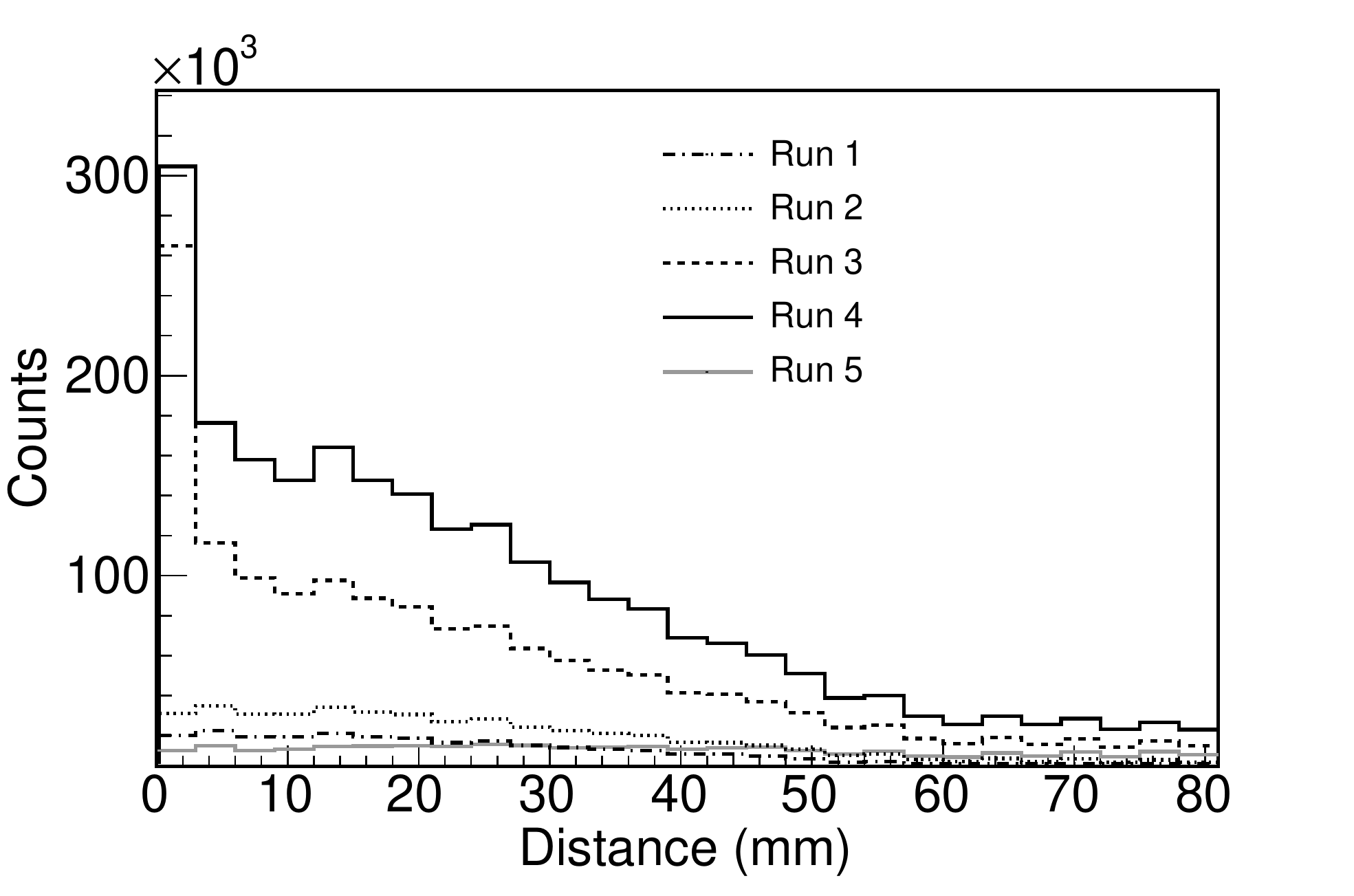}
\caption{ 
Distributions of perpendicular distances from Source 1 to  lines connecting any two hits for the five simulation runs.
Two hits with perpendicular distance $d <d_{\rm max}$ (2.9 mm  $< d_{\rm max} < $  4.1 mm) together with opening angle $\theta > 90^\circ$ are considered  as a coincidence.
The definitions of $d$, $d_{\rm max}$ and $\theta$ are shown in Fig.~\ref{fig:coincidence}. }
\label{fig:h1distany}
\end{center}
\end{figure}

In order to study efficiencies of event selection processes all coincidences with energy depositions higher than 5 keV were selected.  
The data from previously described simulation runs were used with selection conditions as in Section~\ref{sec:selalg}. 
Fig.~\ref{fig:h1distany} shows distributions of perpendicular distances $d$  from Source 1 to lines connecting events with two hits. 
Scuh events for which $d<d_{\rm max}$ (in the first bin) and the opening angle $\theta >90^\circ$ were considered as a coincidence. % as coincidence.

One million events were simulated for each run assuming point-like source geometry.
The percentages of events with coincidence hit pairs $f_{\rm c}$ as well as the number of detected events $f_{\rm d}$ are shown in Table \ref{tb:runs}. 
The percentage of events with coincidence pairs selected for the collinear gamma-ray simulation run is equal to 17.8\%.  
It is slightly smaller than 19.6\% which is the probability to detect both annihilation gamma-rays
as calculated from the detection efficiency of one single 511 keV gamma-ray.
%It is due to the fact that the events can not be selected if the two back-to-back
%gamma-rays deposit energies lower than the threshold during their first interaction; however, both of them can still be detected if the scattered gamma-rays deposit energy higher than the threshold in other i.e. not aligned bars. 
One of the reasons is that the events can not be selected if the two collinear
gamma-rays deposit energies lower than the threshold during their first interaction.
However, both of them can still be detected if the scattered gamma-rays 
deposit energy higher than the threshold in other i.e. not aligned bars. 
Furthermore, the gamma-ray travelling through the side of the detector has lower chance of depositing 
its energies than the other one in the case of collinear gamma-rays.

The percentage of selected events with coincidence hit pairs from the $^{22}$Na run is  equal to 18.2\%.  
The collinear gamma-rays are generated in 90.3\% of all $^{22}$Na decays and 17.8\% of them are found with the valid coincidence pattern.  
Thus, the absolute percentage of selected coincidence events should be equal 
to 16.1\% assuming an absence of gamma-rays from the decay channel with the energy of 1274 keV.
About 11.5\% of all coincidence events selected from $^{22}$Na decays, i.e. 2.1\% of total events, 
are due to extra coincidences between gamma-rays with energies of 511 keV and 1274 keV.
It can be seen in Table~\ref{tb:runs} that only 0.6\%, 1.0\%  and 0.3\% of events from runs 1, 2 and 5 respectively are coincidences with a multiple scattering. 
In conclusion, the introduced coincidence selection method is able to effectively select hits with the collinearity pattern i.e. 
coming from the $^{22}$Na decay while other events are rejected.

\begin{figure}[H]
\begin{center}
 \includegraphics[width=0.49\textwidth]{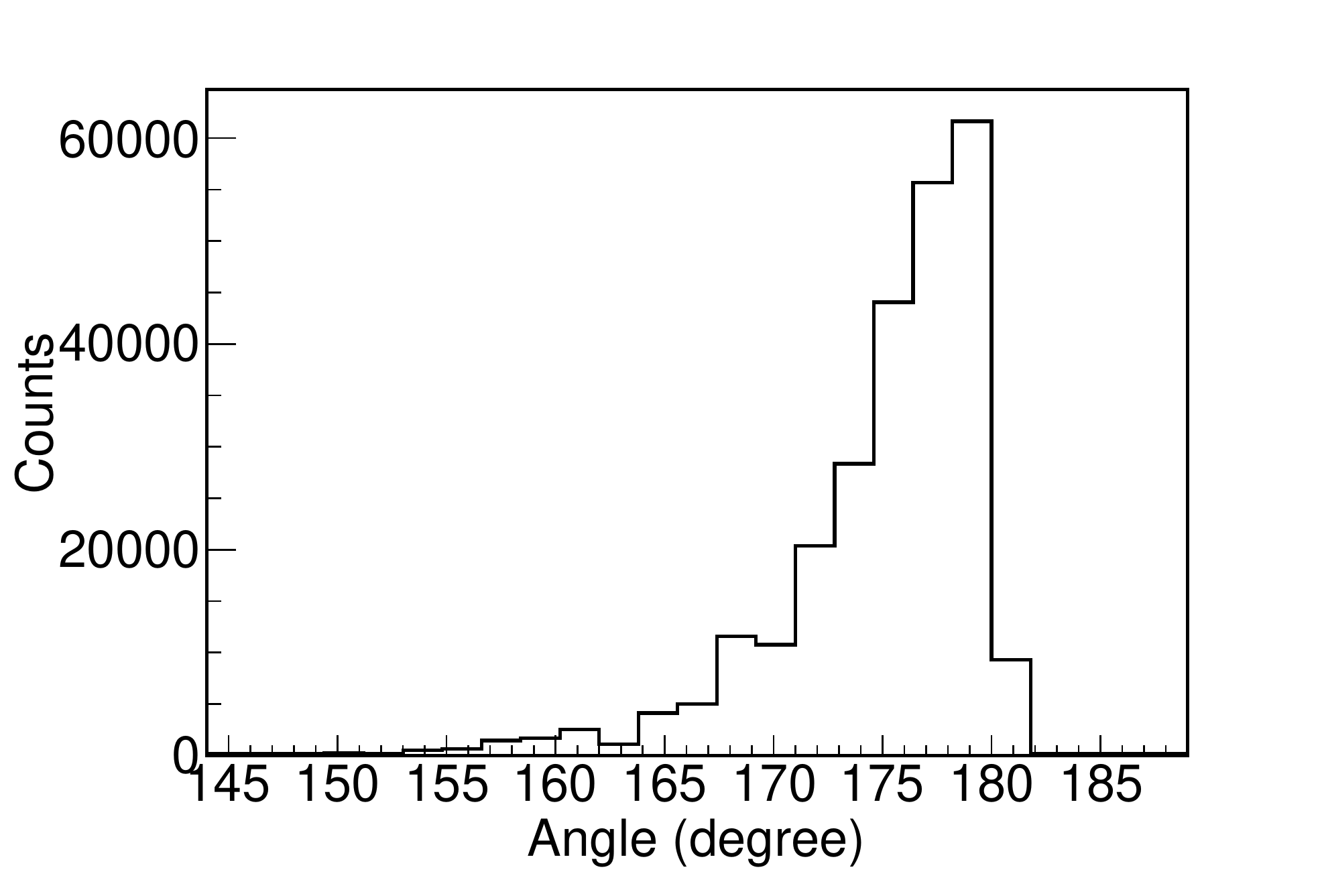}
  \includegraphics[width=0.49\textwidth]{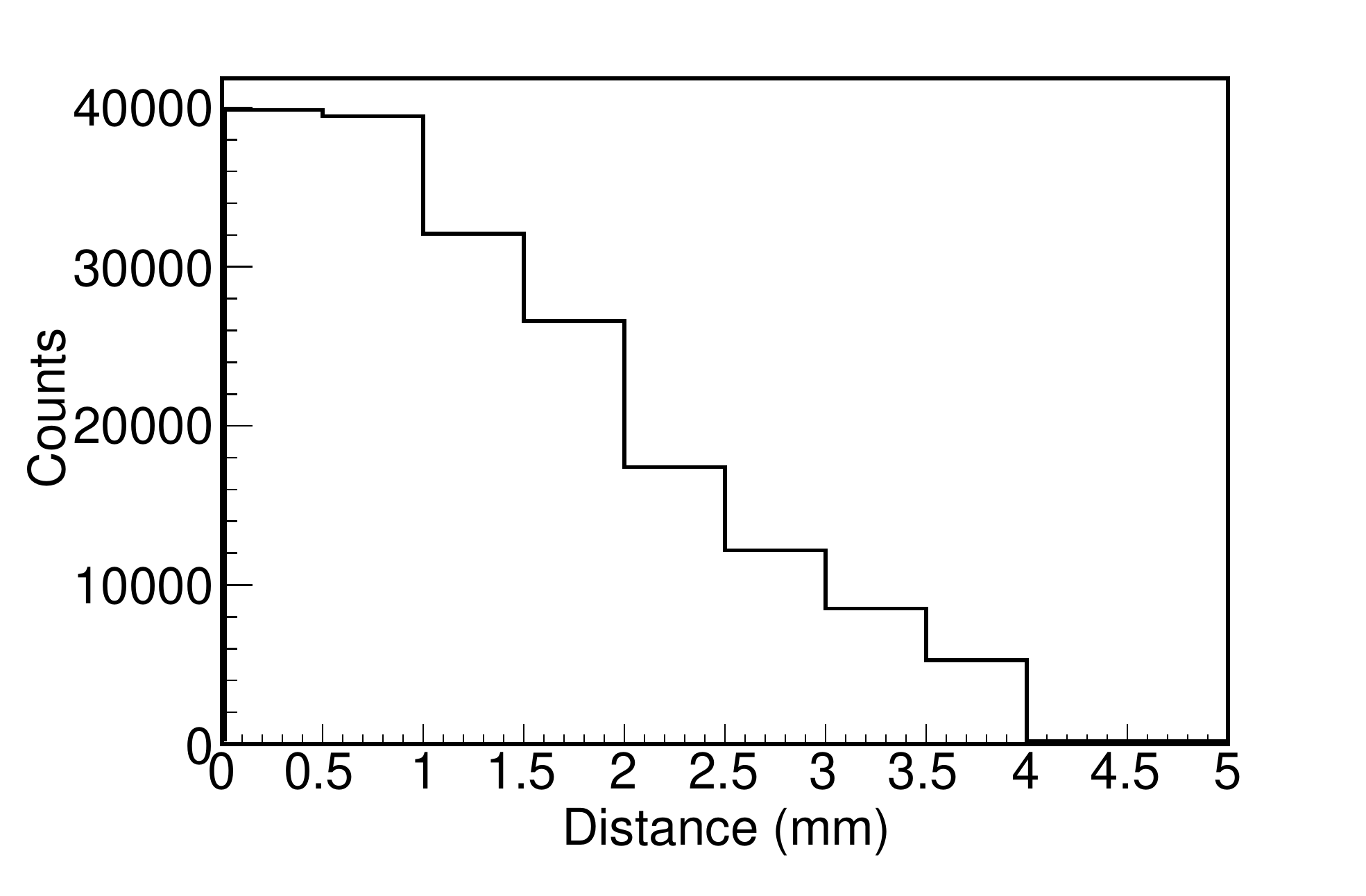}
\caption{
Distributions of opening angles $\theta$ (left) and perpendicular distances $d$ (right) extracted using data from the $^{22}$Na 
simulation run.
}
\label{fig:opening_angle_distance}
\end{center}
\end{figure}
%/opt/geant4/g4work/g4POLAR_201805_forPaper/Na22FinalSim_201804/coincidence/*.pdf

Fig. \ref{fig:opening_angle_distance} shows distributions of the 
opening angles $\theta$ and the perpendicular distances $d$ in the coincidence hits from the $^{22}$Na simulation run (i.e. run 4).
It can be seen that the opening angles $\theta$
are close to  180$^\circ$  while $d$ is less than the upper limit of $d_{\rm max}$ (i.e. 4.1 mm) as expected. 
The distribution of distances between selected hits ($d_{\rm pair}$)  is shown in Fig. \ref{fig:hitdist} (see chapter 7). 
According to the simulation,  96\% of the selected hits have distances smaller than 200 mm.  
Thus, a cut of the distance at the value of 200 mm can be further applied for the coincidence hit selection.
It will further reduce spectral contamination coming e.g. from accidental coincidences for cases with very high backrount rates.

\subsection{Energy spectra}
\label{sec:smearing}
As described in Ref.~\cite{smearing}, the observed energy spectrum $s(h)$
can be computed by convolving the theoretical energy spectrum with the detector response function $f_{\rm MC}(E)$ 
\begin{equation}
 s(h)=\int g(h, E) f_{\rm MC}(E)dE,
 \label{eq:conv}
\end{equation}
where $h$ and $E$ are the energy depositions expressed in ADC channel and keV, $f_{\rm MC}(E)$ is 
the theoretical energy spectrum (obtained e.g. by means of Monte Carlo simulations) and $g(h,E)$ is the response function of the detector. 
It is reasonable to assume that the detector response $g(h,E)$ for a fixed energy $E$ is Gaussian:  
\begin{equation}
g(h,E)=\frac{a}{\sqrt{2 \pi} \sigma_h}\exp\left(-\frac{(h-c E)^2}{2\sigma_h^2} \right),
 \label{eq:response}
\end{equation}
where $a$ is the normalization factor, $c$ the energy conversion factor (in units of ADC channel / keV) and 
$\sigma_h=c \sigma_E = c E \cdot R(E)$. 
$R(E)$ is the energy resolution function described in Ref. \cite{energyres}: 
\begin{equation}
R (E)=  \sqrt{\alpha^2+ \frac{\beta^2} {E} +\frac{\gamma^2}{E^2}}.
\label{eq:res}
\end{equation}
Parameter $\alpha$ takes into account contributions for material inhomogeneities and imperfect light coupling,
$\beta$ contains statistical effects and $\gamma$ describes electronic noise.

\begin{figure}[H]
 \begin{center}
 \includegraphics[width=0.8\textwidth]{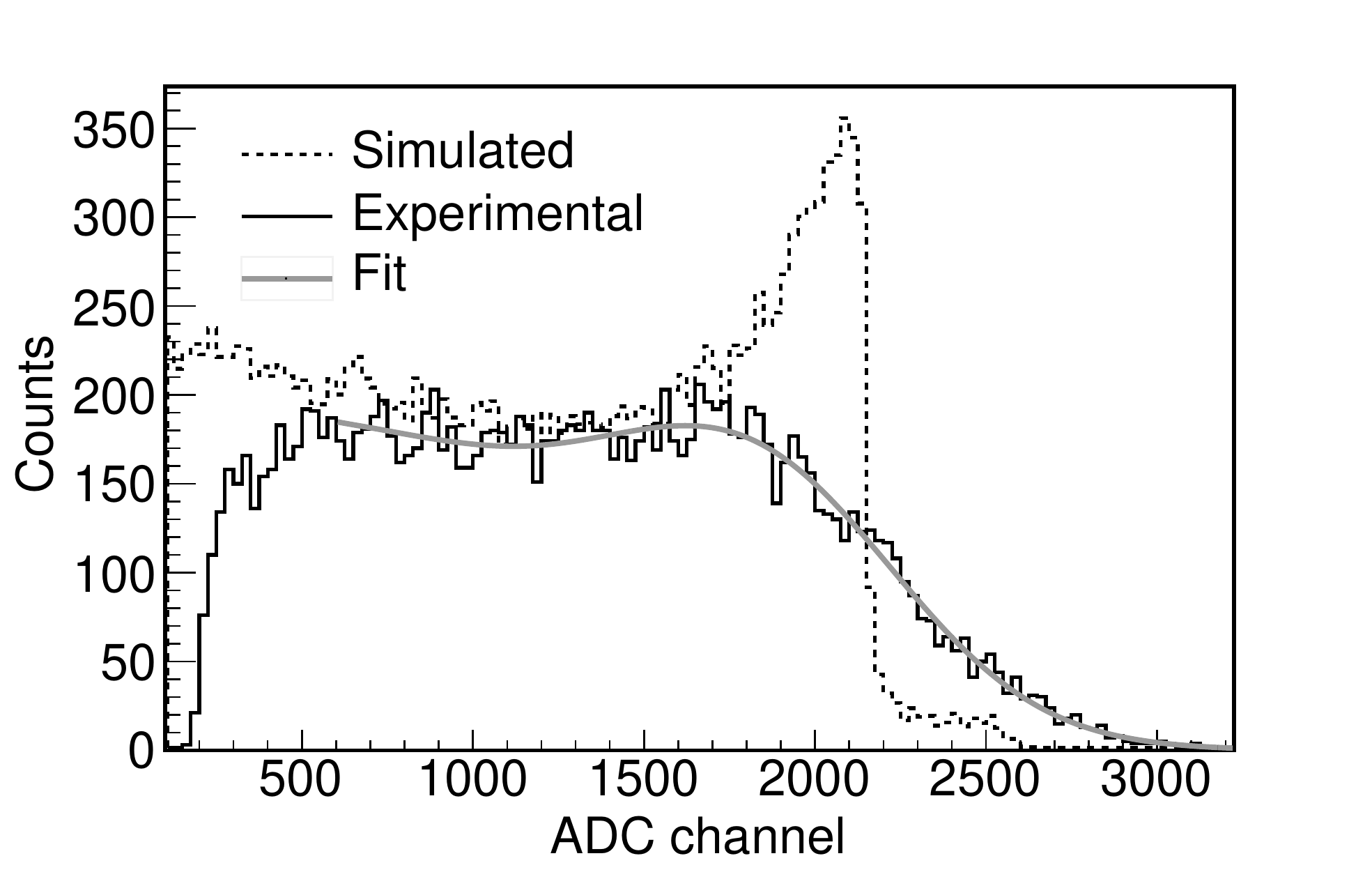}
 \caption{ A typical experimental energy spectrum measured for a 511 keV X-ray beam (solid line) and 
  a fit of the convoluted  spectrum as given in Eq.~(\ref{eq:conv}) (grey line) with $\alpha=0.16$, $\beta=1.0$, $\gamma=3.7$, $c=6.3$ and $a=2.44$.
  The  $\chi^2$ / degrees of freedom (NDF) is equal to 90.9/109. The dotted line shows the Monte Carlo simulated energy spectrum converted to the ADC channel scale.
 %p2=1.5, beta=sqrt(1.5)
 %
 }
 \label{fig:parafit}
 \end{center}
\end{figure}

At the beginning of the study, we selected typical energy spectra (e.g. the black solid line in Fig.~\ref{fig:parafit})
measured for monochromatic gamma-rays at the European Synchrotron Radiation Facility (see Refs.~\cite{silvio,hlastro} for details). 
This way we obtained experimental values of the POLAR energy resolution at different photon energies. 
Corresponding simulation runs were performed with GEANT4 software suite to generate theoretical energy spectra.
Each experimental spectrum was fitted to the convoluted one $s(h)$ using the least squares method \cite{root}.  
Note that  $\alpha$, $\beta$ and $\gamma$ in the energy resolution function as well as $c$ and $a$ 
were free parameters directly determined in the fit. 

The grey line and the dotted line in Fig.~\ref{fig:parafit}  show  the best fit convoluted spectrum $s(h)$   and 
the simulated energy spectrum respectively. 
It is worthwhile to remark that the simulated energy spectrum was scaled along the x-axis by the energy conversion factor $c=6.31$,  determined from the fit. 
The other fit parameters $\alpha$, $\beta$, $\gamma$ and $a$ were equal to 0.16, 1.01, 3.68, 6.31, 2.44 respectively.  
The energy resolutions values at 477.7 keV and 340.7 keV are equal to 16.0\% and 16.2\% respectively.  
They are consistent with the values reported in Refs.~\cite{silvio,zhangxfgain}. 
%~/svn_psi_polar/software/energyResolutionFit# ./spectrumUnfoldingESRF
%root [2] Info in <TCanvas::SaveSource>: C++ Macro file: /home/xiaohl/svn_psi_polar/software/energyResolutionFit/parameterFit.cc.C has been generated

\begin{figure}[H]
\begin{center}
 \includegraphics[width=0.49\textwidth]{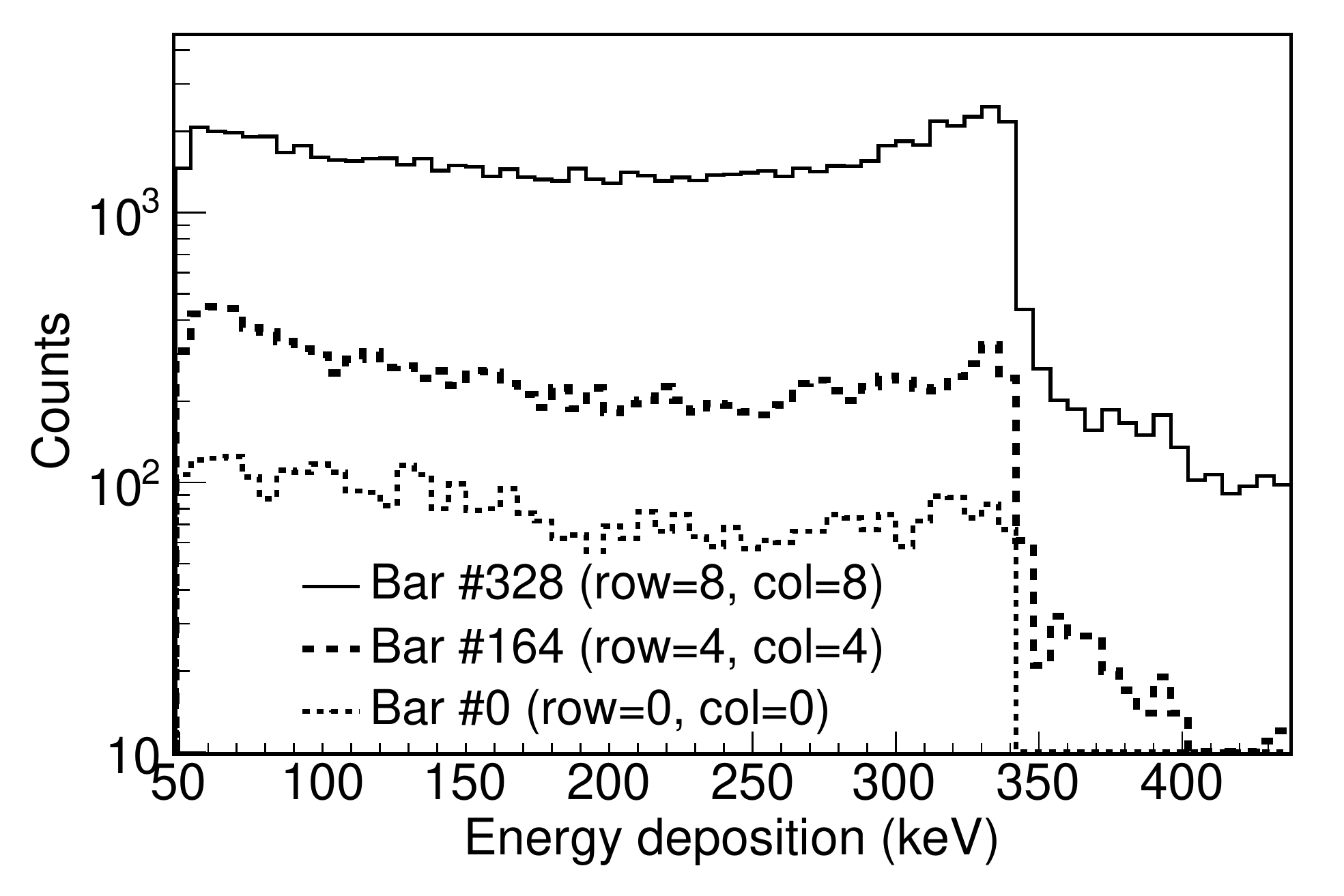}
 \includegraphics[width=0.49\textwidth]{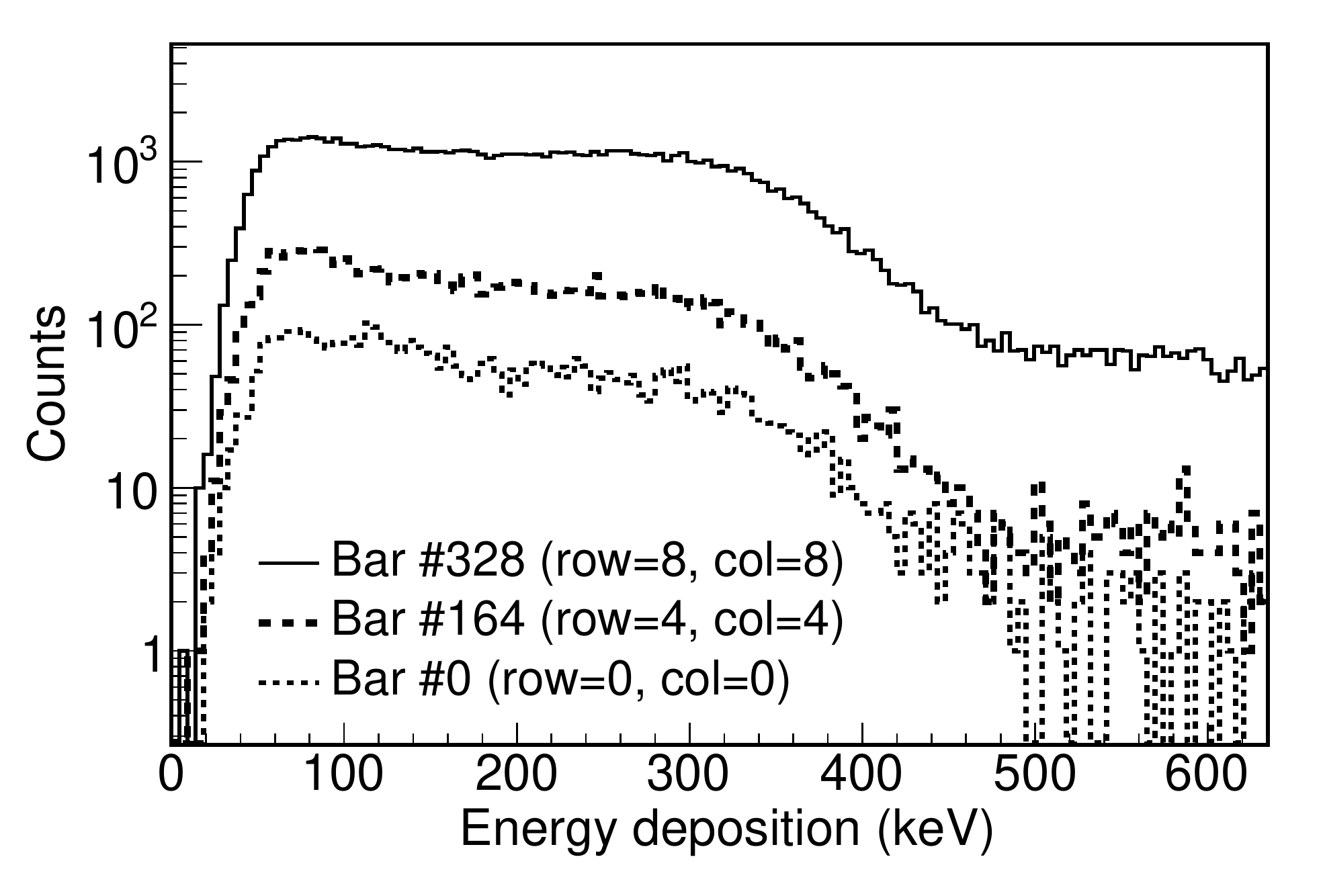}
\caption{Simulated spectra of deposited energy obtained with  selected coincidence hits for 
three typical bars before (left) and after (right) convolving with the detector response function given by Eq.~(\ref{eq:response})
with $c=1, a=1, \alpha=0.16, \beta=1.0$ and  $\gamma=3.7$.
}
\label{fig:energyspectra}
\end{center}
\end{figure}

An extended simulation run with the $^{22}$Na source and 10 million of incident events was performed to generate enough statistics for
energy spectra studies. 
Coincidence hits were selected using the algorithm from Section \ref{sec:selalg} in order to construct the energy spectra.
A typical example for three typical bars is given in the left panel of Fig.~\ref{fig:energyspectra}. 
All spectra were convoluted with the $g(h,E)$ function parametrized as described above.
Three convoluted spectra are shown in the right panel of Fig.~\ref{fig:energyspectra}. 
Note that the energy conversion factor $c$ was set to 1. 
The Compton edges at the value of 340.7 keV can be still clearly seen for all the bars.
The counts on the right side of Compton edges are due to multiple scattering of gamma-rays inside
the bars or extra coincidences between gamma-rays of 511 keV and 1274 keV.
They can be reduced by applying a high energy cut in the hit selection procedure. 
%The energy conversion factors can be obtained by fitting the Compton edges with a step-like function described in Ref.~\cite{silvio}.
%~/svn_psi_polar/software/energyResolutionFit

\subsection{Space backgrounds}
\begin{figure}[H]
\begin{center}
 \includegraphics[width=0.76\textwidth]{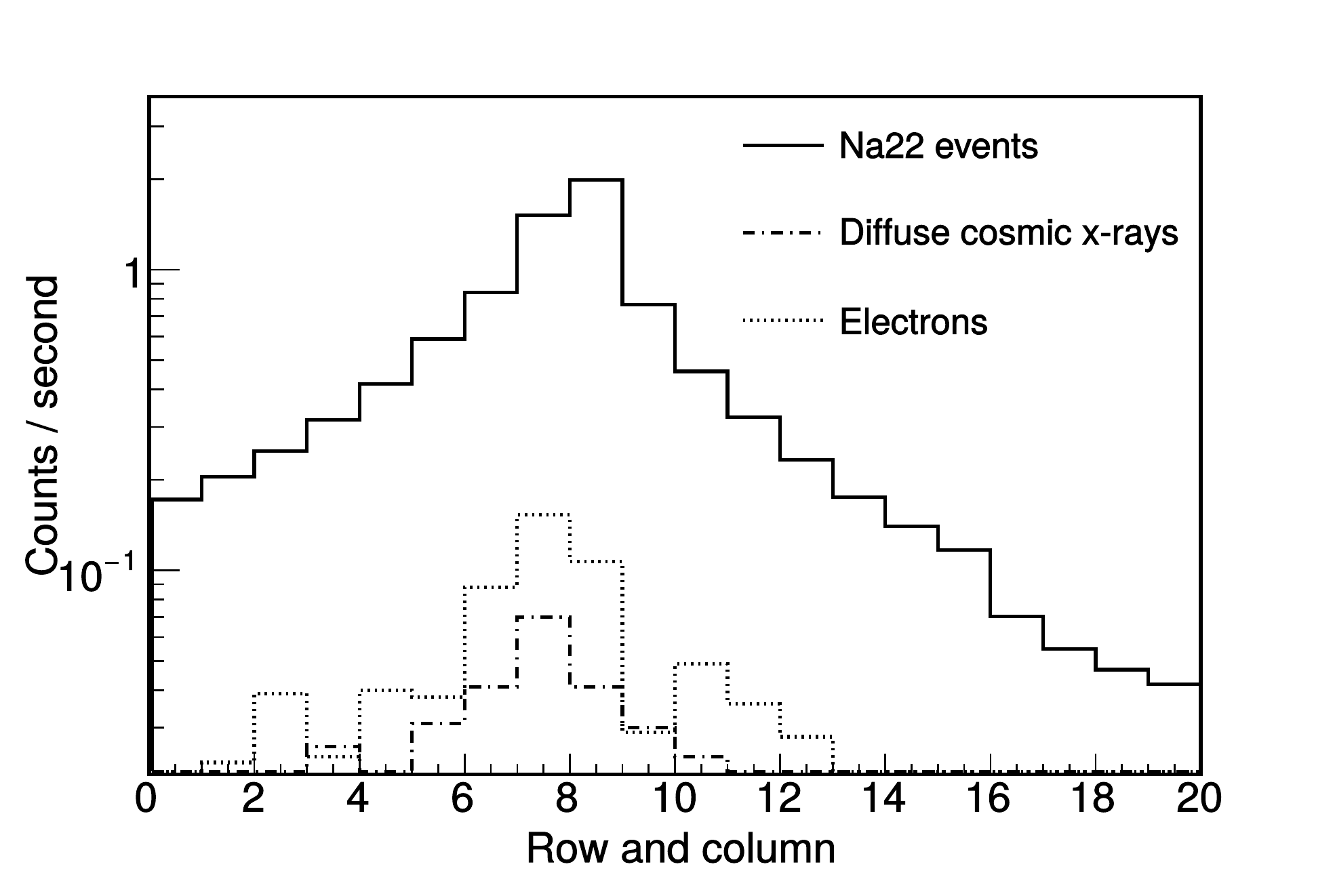}
\caption{ Simulated coincidence hit rates of the diagonal bars 
for the $^{22}$Na source with the activity of 200 Bq, space diffuse cosmic X-rays  and  electron background. 
}
\label{fig:countrate}
\end{center}
\end{figure}

According to simulations described in Ref. \cite{estella} the main sources of the POLAR background in space apart the South Atlantic Anomaly (SAA) are diffuse cosmic X-rays, electrons and positrons.  
Other sources such as neutrons and primary cosmic rays are either negligible or 
easily rejected by the trigger system.
Our simulations showed that positrons can not contaminate Compton edges in calibration spectra as most of them are stopped by the POLAR 
enclosure and the energy spectrum after their annihilation is roughly the same as from calibration sources \cite{estella}.  
Thus, simulations of the other two background sources, i.e. diffuse cosmic X-ray and electrons were performed. 
Energy spectra for both background sources were taken from Ref. \cite{estella}. 
Incoming particles were generated on the surface of a sphere with a radius of 30 cm. 
The coincidence hit selection method as described above was applied to the simulated data.
Fig.~\ref{fig:countrate} shows the number of coincidences per second in POLAR diagonal bars  
obtained for: the 200 Bq strong $^{22}$Na source placed at the Source 1 position,
the diffuse cosmic X-ray background and the electron background.  
Threshold values for the hit selection were also equal to 5 keV.
It can be seen that the coincidence hit rates of the background are much lower than for the calibration signals.  
Furthermore, most of the coincidence hits selected from the background events have rather low energy depositions. 
Only the hits with energy depositions larger than about 200 keV could contaminate the Compton edges. 
The signal to noise ratio of each bar can be improved even further if higher threshold values are applied.
\subsection{Activity of calibration sources}
Duration of POLAR mission extends up to three years. 
Activities of the calibration sources after 3 years will be reduced to 45\% of their initial values. 
According to Fig.~\ref{fig:countrate}, the coincidence hit rates from calibration sources will be above the background rates if the source activity is higher than about 20 Bq. 
Obviously, much longer calibration time periods will be required  in order to obtain similar statistical uncertainties.
%Detector performances, e.g. temperature effect, ageing, can be obtained from the extensive calibrations at the beginning of the mission.
Using sources with higher activity shortens the calibration time but also increases dead time and real background rates 
by worsening the signal to noise ratio of the GRB polarisation measurements.  
The sources of higher activities  will also need larger data bandwidth from the space-lab.
Based on above considerations, the source activity of about or slightly less than 250 Bq was proposed. 
%\st{
%Simulations showed that statistical errors in the energy conversion factors for the outermost bars
%obtained using 100 Bq strong calibration sources are of the order of 1\% for 24 hours long data taking.}

\section{Laboratory verification of the method}
\label{sec:verification}
The first experimental test of the in-flight calibration method was carried out with the
POLAR flight model spare (FMS).  
Four small $^{22}$Na sources with a total activity of about 1000 Bq were prepared for the experiment.  
The dimensions of each source  were  approximately  3 $\times$ 3 $\times$ 0.3 mm$^3$. 
Each source was glued to a L-shape plastic support. 
Finally the supports with their sources were glued to outer edges of four POLAR corner modules as shown Fig.~\ref{fig:na22loc}. 
The data acquisition was started when the temperature of POLAR after switching its power on was sufficiently stable. 
The high voltage values for all the modules were set to 650 V. 
The average total trigger rate per module was around 70 Hz. 
In order to cross-check the results, another run was performed with
a 10 $\mu$Ci  $^{137}$Cs source placed at a distance of  40 cm from the top surface of the POLAR FMS.

\begin{figure}[H]
\begin{center}
 \includegraphics[width=0.6\textwidth]{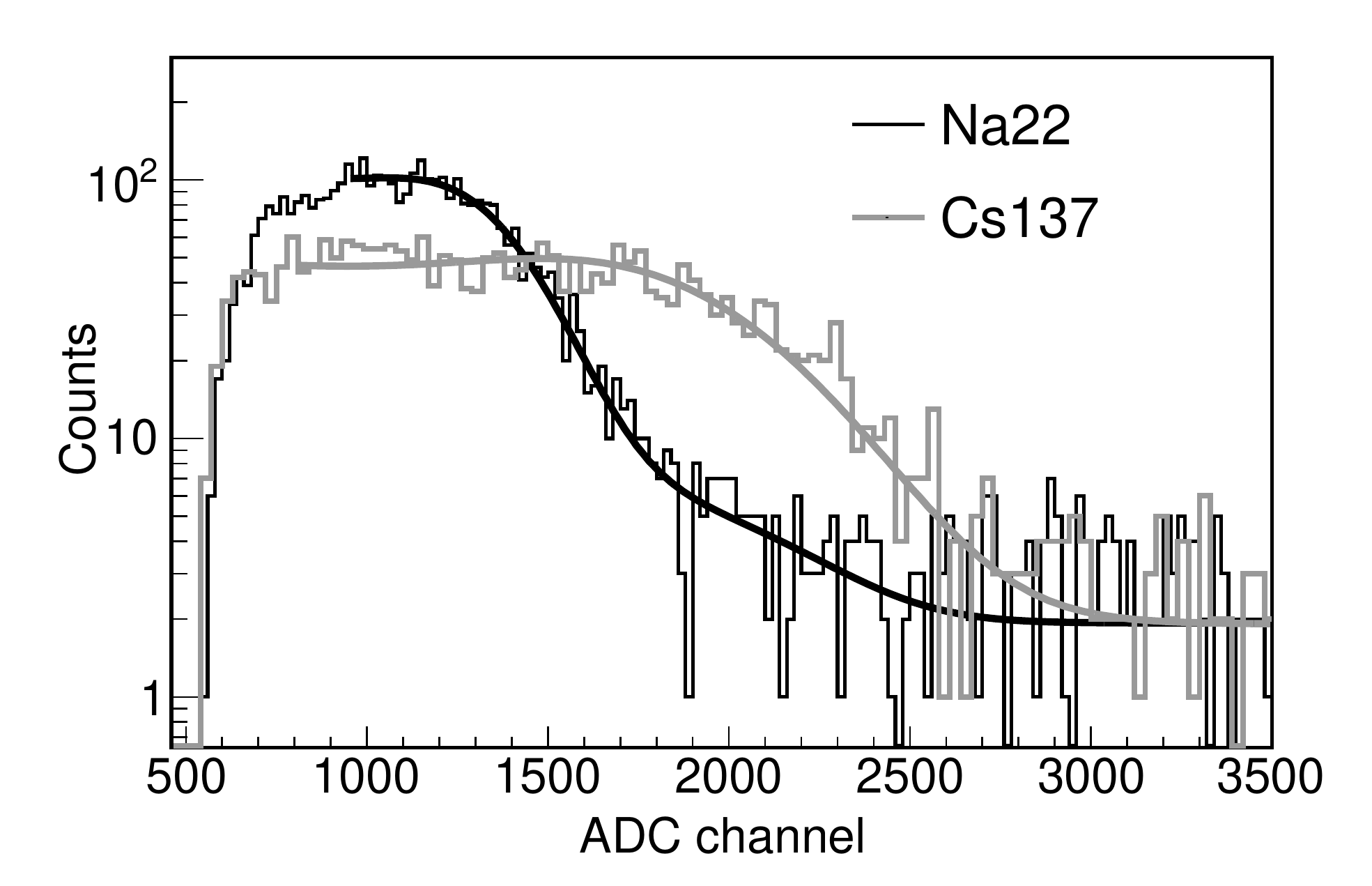}
\caption{
	Experimental energy spectra in the channel 13 of module 25 taken using $^{22}$Na and $^{137}$Cs sources.
The black line represents the spectrum of the coincidence hits selected for the internal $^{22}$Na calibration sources.  
The grey line shows the spectrum measured with the external $^{137}$Cs source. 
The best fit spectra using the method described in Section \ref{sec:smearing} are also shown.
The Compton edge position determined from the fits are equal to $1367\pm11$ and $1964 \pm 16$ AD channel.
$\chi^2$ / degrees of freedom of the fits are equal to 141.2/117 and 113.9/91 respectively.
}
\label{fig:ce_examples}
\end{center}
\end{figure} 

\begin{figure}[H]
\begin{center}
 \includegraphics[width=0.7\textwidth]{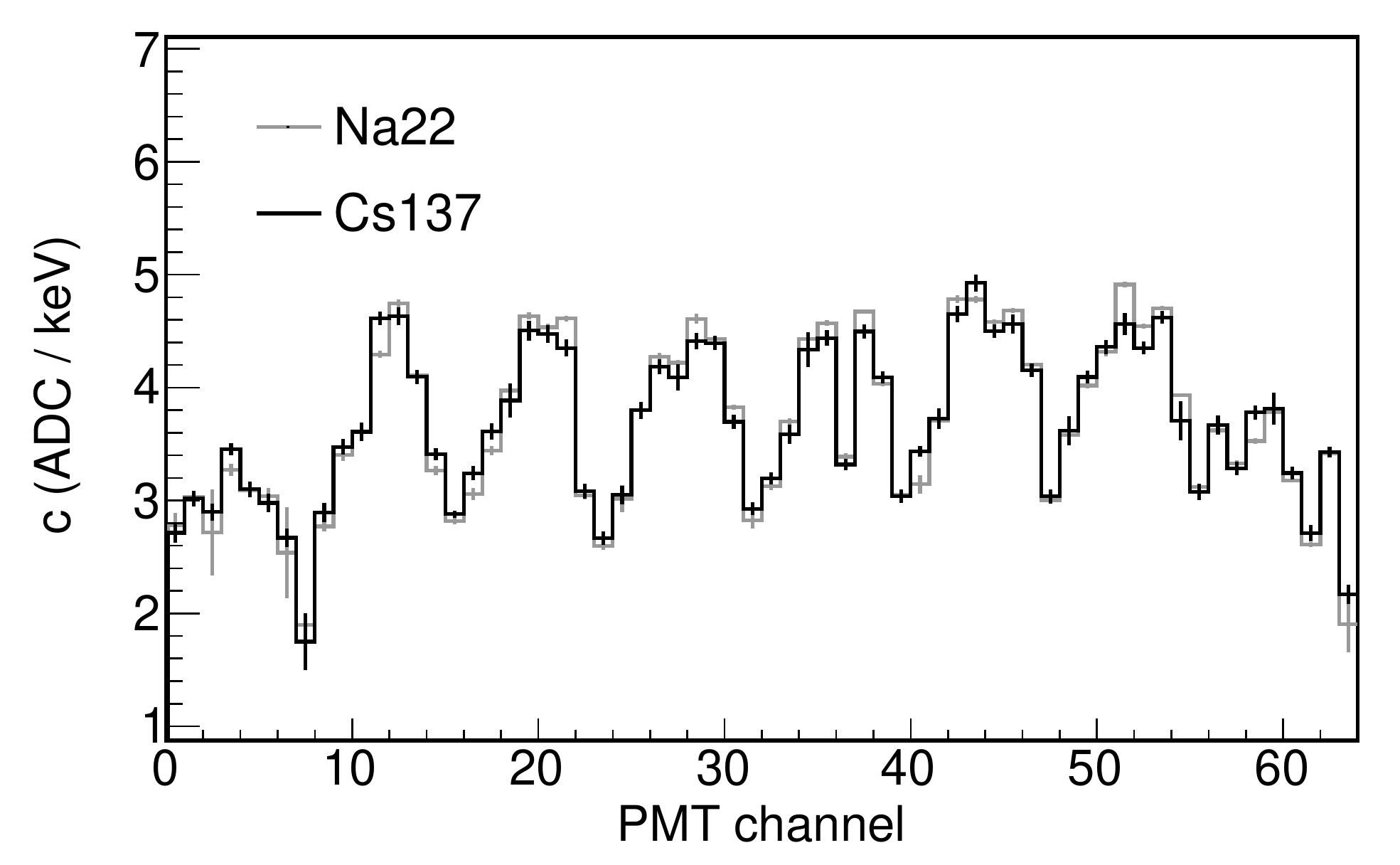}
\caption{
Comparison between energy conversion factors of module 25 obtained with an internal $^{22}$Na source and an external  $^{137}$Cs source.
The figure also indicates the response non-uniformity within a module. (The numbering convention of POLAR channels within a module is adopted from the  Hamamatsu H8500 MAPMT \cite{h8500}.) 
%Chi2=218, NDF=63.
}
\label{fig:na22cs137}
\end{center}
\end{figure}

The data taken from above runs was processed in several steps. 
Firstly, it was decoded and re-written using the ROOT file format \cite{root}. 
Pedestal events taken periodically during data acquisition were 
pre-selected to calculate the pedestal positions (i.e. baseline of signals) for each input channel. 
Subsequently, for each detected event the pedestal values were subtracted. 
In the next step the common noise values (i.e., common shifts of baseline)
were also subtracted from the amplitudes of detected events.
%Note that the events with large common-mode noise values were excluded. 
Finally, the hits belonging to the same physical event were merged 
using hit pattern information from the trigger packets and timestamps values from the module science packet. 
The coincidence hits were selected from the 
data taken with the internal calibration sources by using the method as described in Section \ref{sec:selalg}. 
The value of $d_{\rm max}$ calculated for each hit pair was increased by 3 mm 
taking into account the size of the sources.
As an example, Fig.~\ref{fig:ce_examples} shows the energy spectrum of the 
selected coincidence hits  for channel 13.   The energy spectrum measured with the external $^{137}$Cs sources is also shown.
They were fitted to the convoluted energy spectra as given by Eq.~(\ref{eq:conv}).  
The spectra resulted from the fit are also shown in Fig.~\ref{fig:ce_examples}.  
GEANT4 software suite was used to generate simulated energy spectra for each channel.
%The same procedure described in Section \ref{sec:smearing} was used to determine the energy conversion factors. 
The energy conversion factors determined from the fits are $4.02\pm+0.03$ and $4.11\pm0.05$.
The same procedure was applied for all POLAR FMS channels.  
Fig.~\ref{fig:na22cs137} shows a comparison of the energy conversion factors for all channels in module 25.
A generally good agreement can be seen for energy conversion
factor values determined with two different sources and exposure geometries. 

\section{In-flight Calibration of POLAR}
\subsection{Calibration sources and calibration runs}
\label{sec:inflight}
Four newly produced $^{22}$Na sources were prepared for the POLAR flight model (FM).
Each of them was glued between two L-shape copper foils with a thickness of 0.5 mm each.
The  dimensions of the foil were about $2\times 3$ mm$^2$ for the short side and about $4\times 3$ mm$^2$ for the long one. 
The foils  as well as the glue prevent positrons from escaping. 
The total activity of all sources measured in January 2016  using the germanium detector was equal to 520 $\pm$ 52 Bq (10\% systematic error). 
Activities of single sources were between 100 Bq and 200 Bq.
The foils with the  sources were installed inside of POLAR FM shortly before shipment to the launch site. 
They were glued to the edges of four corner modules. 
The locations of the sources are a few mm away from the module edges pointing to the instrument centre as shown in Fig.~\ref{fig:na22loc}.
More details on the fabrication of the sources and their exact positions in FM are given in Ref.~\cite{polarfm}.

\begin{figure}[H]
\begin{center}
 \includegraphics[width=0.7\textwidth]{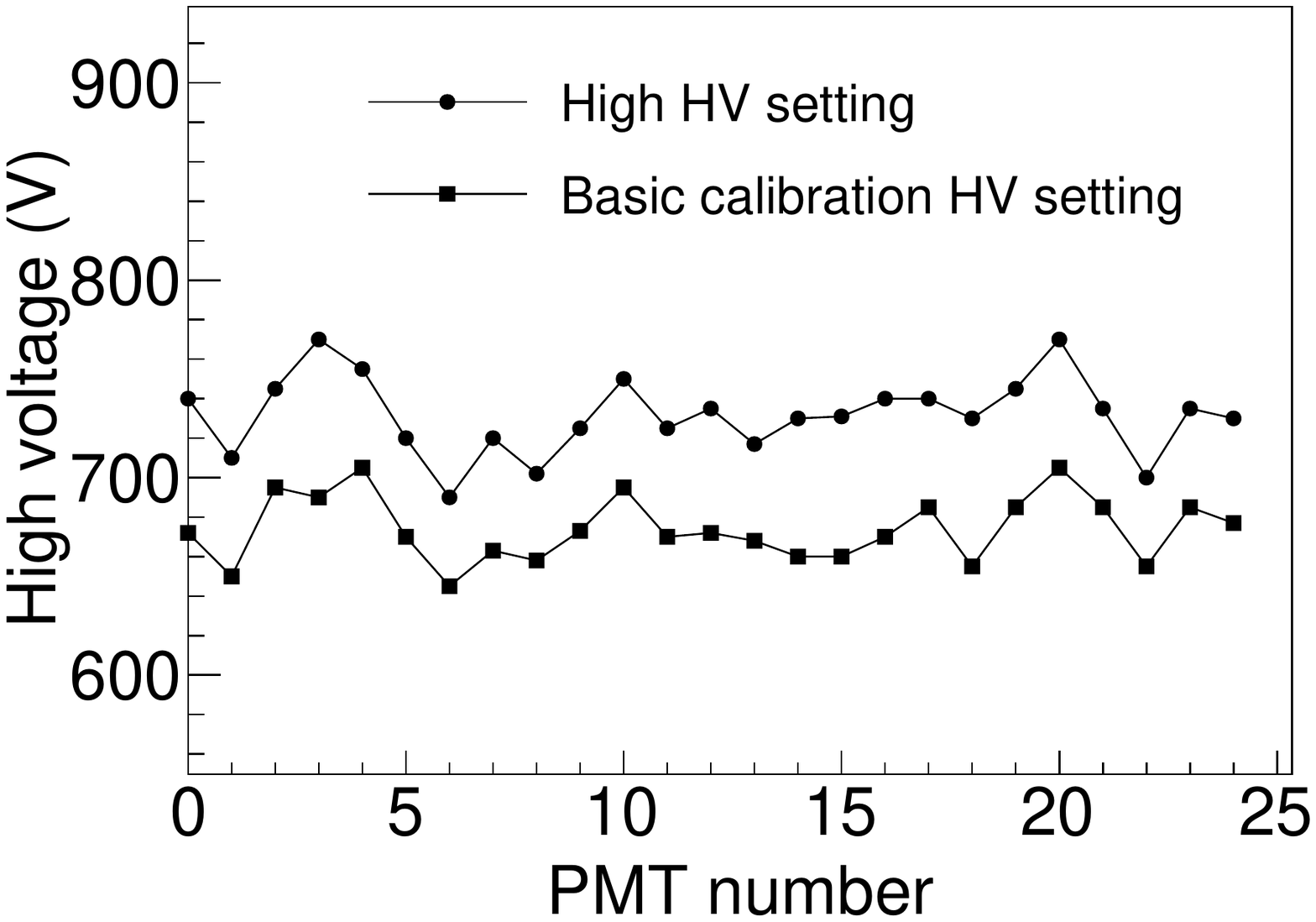}
\caption{
Two settings with HV values used for FM in space. 
Basic in-flight calibration settings are represented by squares.
Full circles show settings optimized for GRB observations.
}
\label{fig:na22hv}
\end{center}
\end{figure} 
Shortly before the launch several HV and threshold settings were prepared and optimized for both in-flight calibrations
and nominal observations of GRBs. 
The optimization method as well as laboratory verification procedures can be found in Ref.~\cite{icrc1}.
HV values for the basic in-flight calibration setting are shown in Fig.~\ref{fig:na22hv}. 
They assure that Compton edges for all POLAR channels are seen in the spectra of coincidence hits.
During POLAR operation in space several in-flight calibration runs with above settings have already been performed.

\subsection{Calibration data processing chain}
A dedicated data centre  was established at Paul Scherrer Institute in order to store and process all POLAR space data.
The raw datasets arriving at the centre are firstly preprocessed, decoded and converted into the level-0 data.
One generally uses the ROOT file format. 
Data coming from different POLAR modules as well as housekeeping packets are stored separately. 
In order to form the level-1 data set all pedestals and common mode noise values are subtracted for each physical event. 
In the next step of data processing all hits belonging to the same physical event recorded at different modules are merged. 
Additionally the merged events contain all relevant housekeeping data such as module temperatures and HV values.  
They are written to the new data files making the level-2 dataset. 
After data merging, the coincidence algorithm is applied to the level-2 dataset to extract the in-flight calibration events. 
The maximum allowed perpendicular distance $d_{\rm max}$ for each hit pair is increased by
3 mm to include both the source size and the uncertainties of its location.
Events with energy depositions above the ADC range or with too many channels triggered, e.g. caused by cosmic-rays, are excluded.
Moreover, all hits for which the neighbouring bars have higher energy depositions are also excluded in order 
to filter out excessive crosstalk or small-angle Compton scatterings. 
All events selected with the help of above criteria are written to new data files assigned as level-2B.

Compared to the level 2 data, all events in the level-2B dataset contain extra information such as  
e.g. positions of two selected bars, their distances to the calibration sources,
the perpendicular distance and the opening angle.  
The selection steps leading to this dataset are performed using an automated 
data processing chain at PSI POLAR data centre.  
Further description of the POLAR data centre is given in Ref.~\cite{icrc2}.

\subsection{Event selection}
\begin{figure}[H]
\begin{center}
 \includegraphics[width=0.42\textwidth]{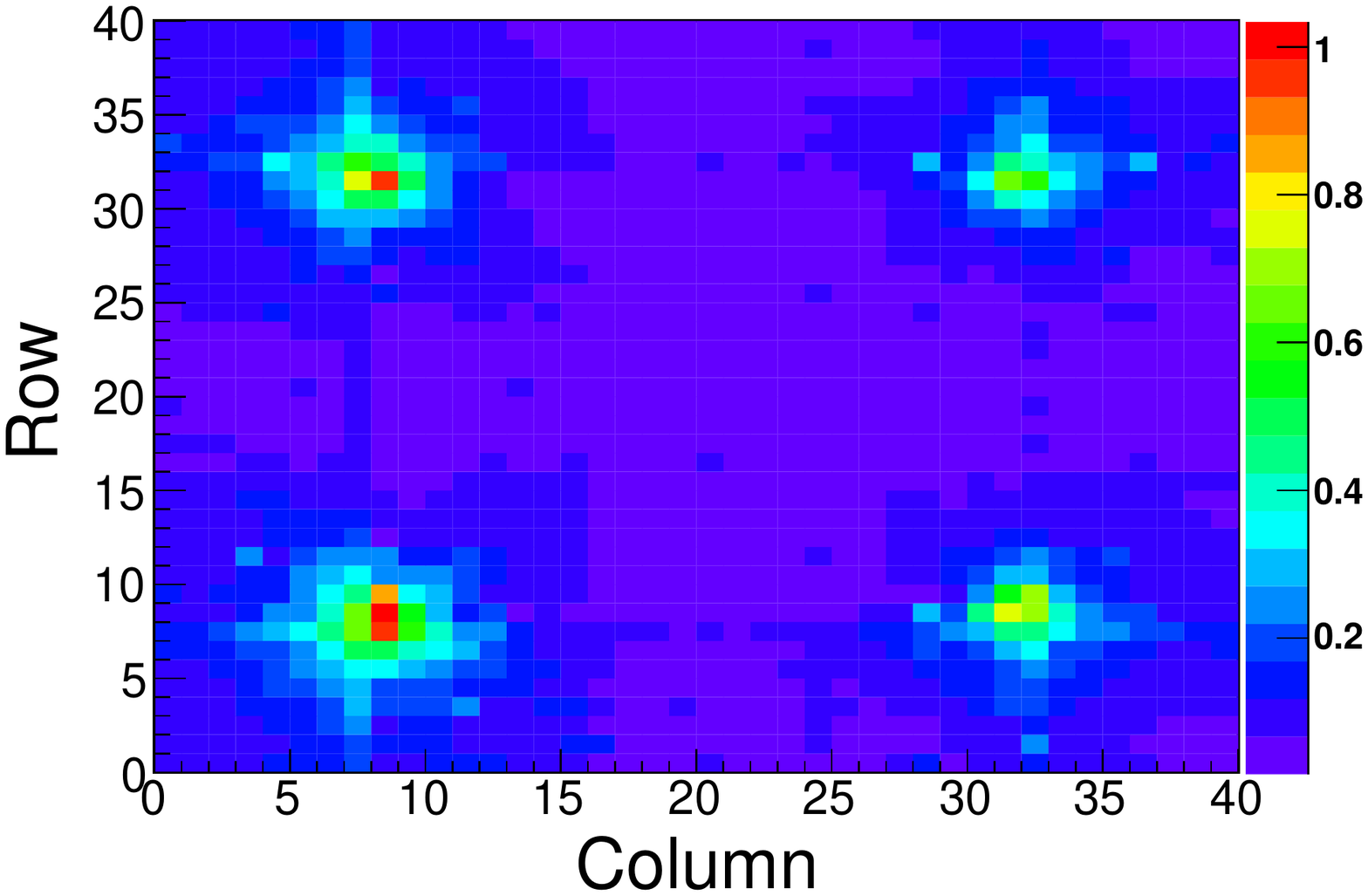}
 \includegraphics[width=0.42\textwidth]{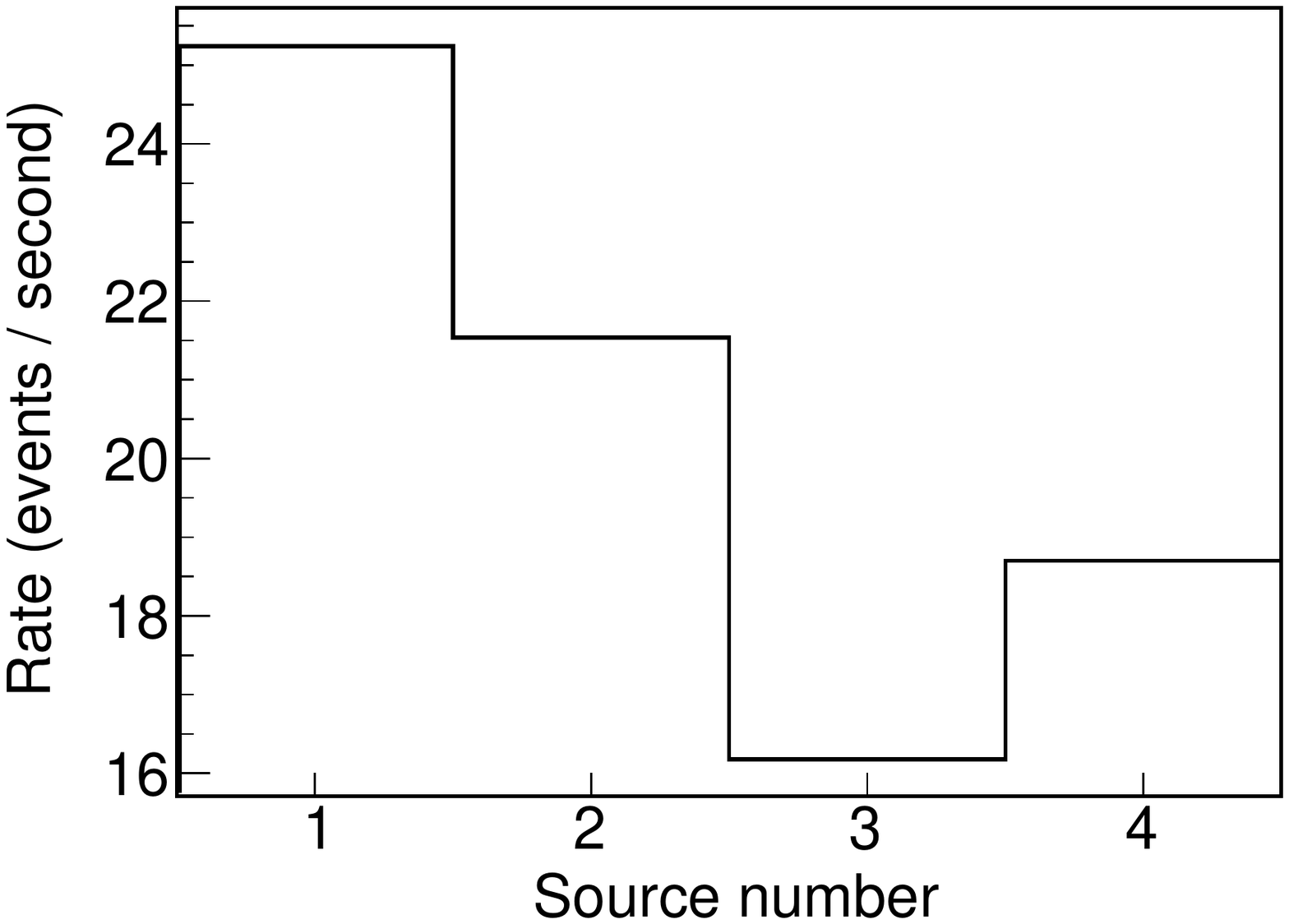}
\caption{
Map of coincidence rates (hit/second) for four in-flight calibration sources  
(left)  and the total coincidence event rate selected  for each calibration source (right).
(The source numbering convention is shown in Fig.~\ref{fig:na22loc}.)
}
\label{fig:hitsel}
\end{center}

\end{figure}

\begin{figure}[H]
\begin{center}
 \includegraphics[width=0.65\textwidth]{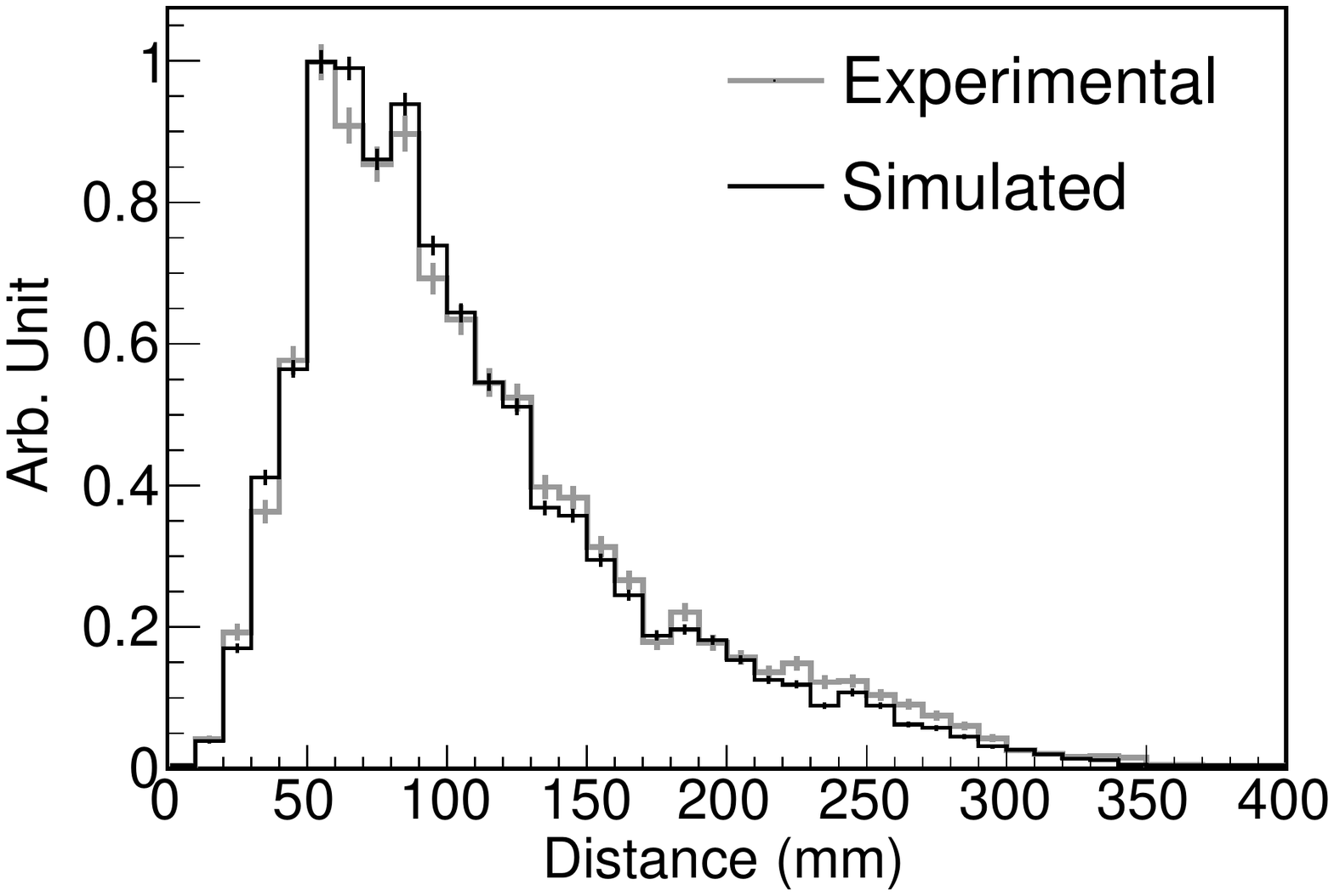}
\caption{
Comparison between simulated and experimental distributions of distances between 
the centres of the two bars with coincidence hits (i.e. $d_{\rm pair}$  - see in Fig.~\ref{fig:coincidence}).
%The mean distance are 107 and 110 mm respectively. %The data are presented for source 1. 
}
\label{fig:hitdist}
\end{center}
\end{figure}
%Chi2 = 83.9, Prob = 1.53119e-06, NDF = 37, igood = 3
%root [3] Info in <TCanvas::SaveSource>: C++ Macro file: /home/xiaohl/svn_psi_polar/software/SimDataCalibration/draw_hexp_bardist_npairsAll_Compare.C has been generated

The left panel in Fig.~\ref{fig:hitsel} shows the 2D map with coincidence rates constructed using 
the in-flight calibration data taken on November 19th, 2016. 
The rate values attributed to each calibration source are shown in the right panel. 
The total event rate for the in-flight calibration data taking 
was equal to 80 $\pm$ 9 Hz corresponding to the summed activity of $^{22}$Na sources equal to 460 Bq.  
Based on measurements from January of 2016 the remaining activity of the sources was equal to 420 $\pm$ 42 Bq.
The mean threshold of POLAR channels was about 22 keV. According to our simulations, 
the percentage of events with coincidence hit pairs was equal to about 17\% at the mean threshold.
It should be noted that the maximum allowed perpendicular distance $d_{\rm max}$ for hit select selection was also increased by 3 mm in the simulations.
Based on above conditions the calculated number of detected $^{22}$Na events was equal to 71 $\pm$ 7 Hz.
%computed with selection criteria
%for coincidence hit pairs as described in section \ref{sec:selection}. 
Thus, the number of calculated event rate and measured during in-flight calibration runs are in good agreement.

\begin{figure}[H]
\begin{center}
 \includegraphics[width=0.6\textwidth]{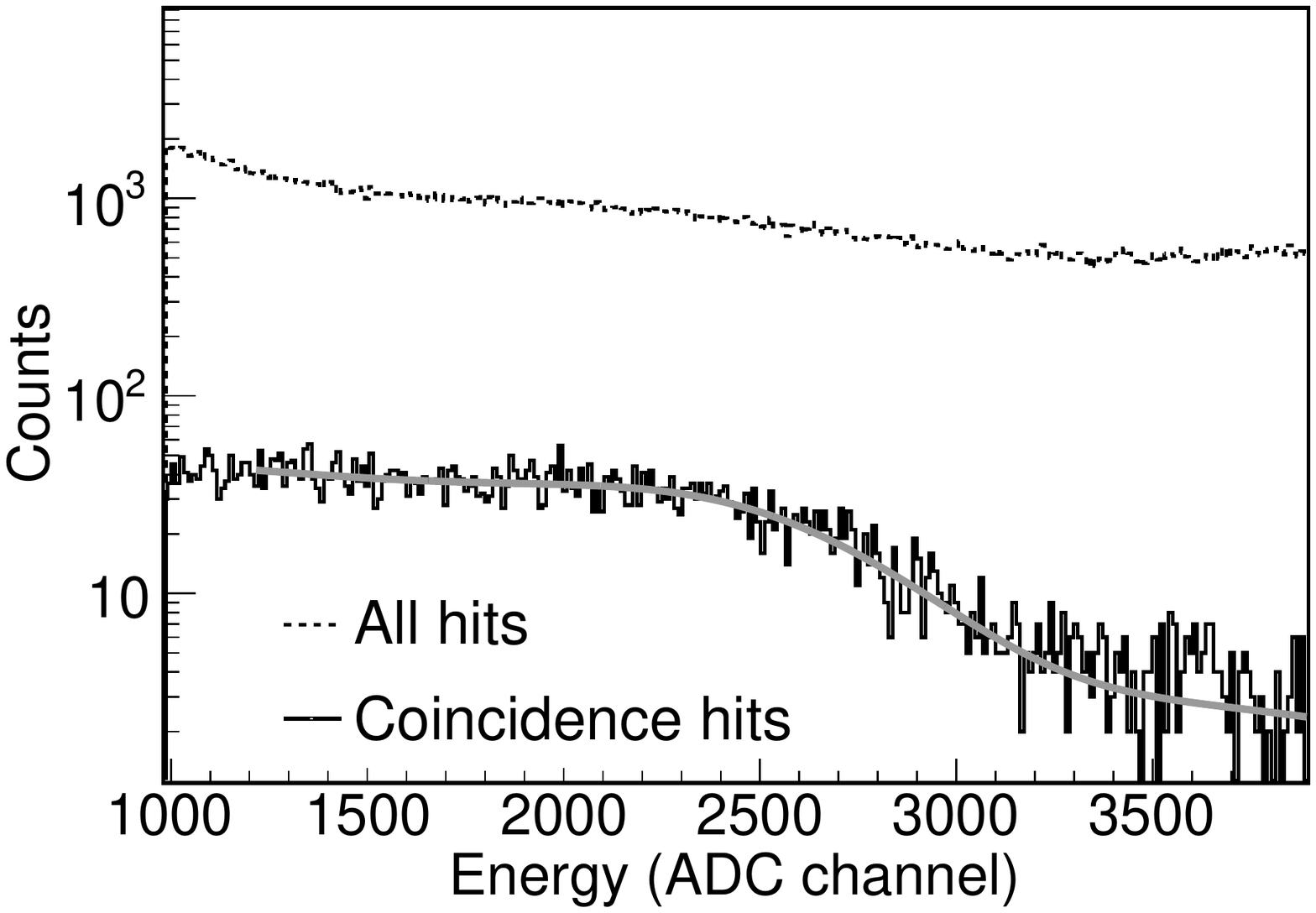}
\caption{Raw and coincidence hits energy spectra from POLAR channel 2 measured during calibration runs on November 19th, 2016.
Coincidence hits selected for all four Na22 sources were used. A fit of the Compton edge based on a simulated spectrum with energy resolution smearing
is also shown.  The Compton edge position from the fit is 2577 $\pm$ 17 ADC channels, $\chi^2$ / degrees of freedom equals to 267.6/249,
the energy resolution $\sigma_E/E$ at 340.7 keV calculated using Eq.~(\ref{eq:res})  is equal to 13.3\%.  
}
\label{fig:ch1spectra}
\end{center}
\end{figure}

The distribution of distances between the centres of the two bars with the selected coincidence hits (i.e. $d_{\rm pair}$  - see in Fig.~\ref{fig:coincidence}) is 
shown in Fig.~\ref{fig:hitdist}. 
It presents a good agreement with the simulated distribution which is also shown in the same figure.
Fig.~\ref{fig:ch1spectra} shows an example of the raw energy spectrum and the spectrum of the coincidence hits for the POLAR channel No. 2. 
This channel is placed in the outermost layer. 
As it can be seen,  there is no Compton edge feature in the raw energy spectrum
but it can be clearly seen in the spectrum of the coincidence hits.  
It shows great enhancement in the signal to noise ratio obtained after applying the coincidence selection criteria.

%/data2/polardata/calibration/Na22/bank20_201708/channel2spectra_Nov19.pdf

\subsection{Energy conversion factors}

\begin{figure}[H]
\begin{center}

 \includegraphics[width=0.9\textwidth]{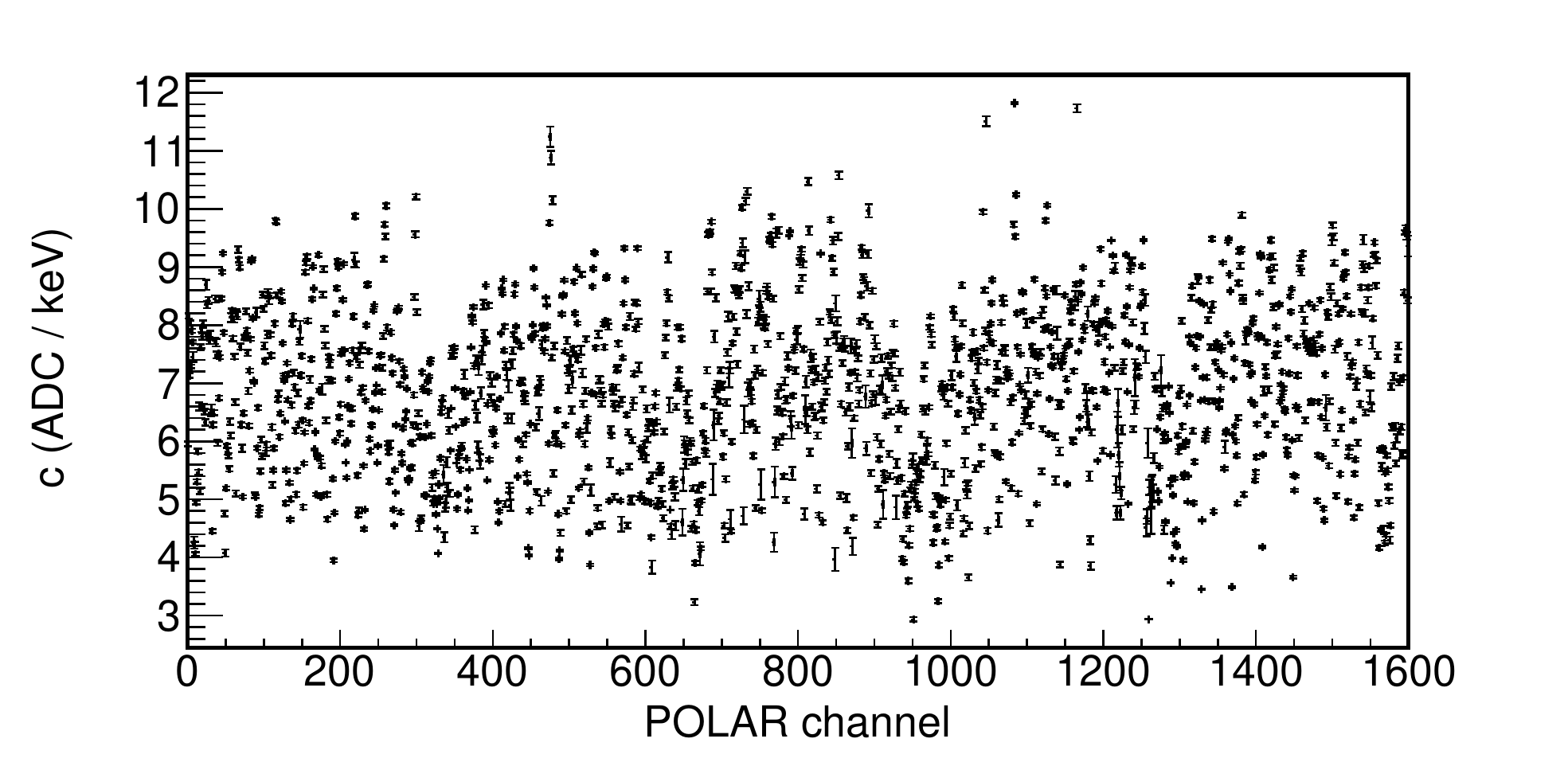}
\caption{Energy conversion factors for all 1600 POLAR channels obtained 
using data taken in the period from September 30th to October 4th, 2016.
The method used is described in Section \ref{sec:smearing} while the HV settings are shown in Fig.~\ref{fig:na22hv}. }
\label{fig:energyfactors}
\end{center}
\end{figure}
%/home/xiaohl/svn_psi_polar/software/energyResolutionFit/unfolding/bank20_ce1600_Oct2016.pdf

\begin{figure}[H]
\begin{center}
 \includegraphics[width=0.6\textwidth]{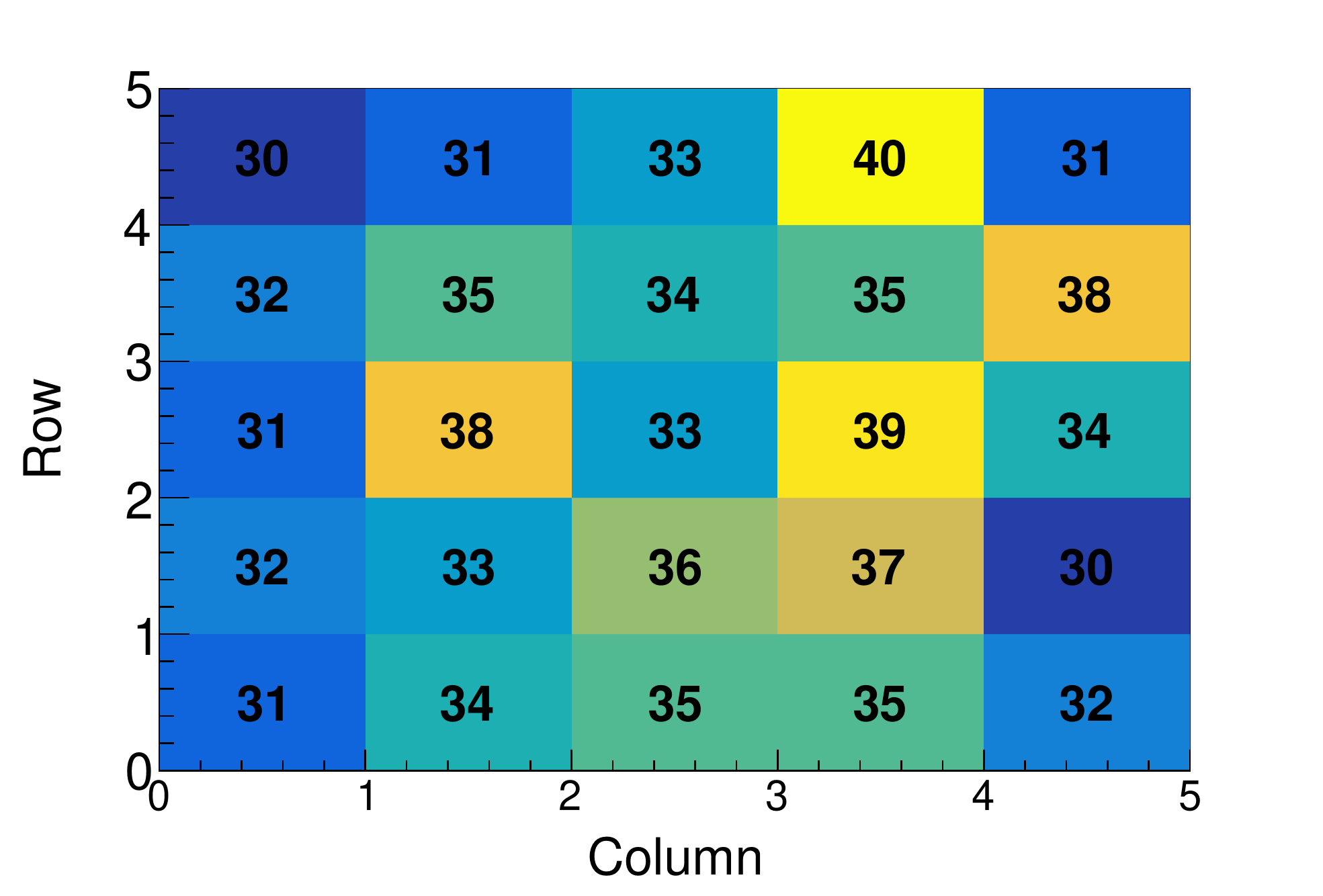}
\caption{Mean temperatures of all 25 POLAR modules (in units of $^\circ$C) measured between September 30th and October 4th, 2016. }
\label{fig:pmtemp}
\end{center}
\end{figure}

The in-flight calibration data taken between September 30th  and October 4th, 2016 were selected to 
study and validate the energy conversion factors.
All coincidence hits in the above time window were used to prepare calibration energy spectra. 
For each bar the Compton edge position in its energy spectrum  
was fitted with the simulated energy spectrum. The energy resolution smearing was based on Eq.~(\ref{eq:conv}). 
The determined energy conversion factors for all 1600 POLAR channels are shown in Fig.~\ref{fig:energyfactors}. 
The mean conversion factor is equal to 6.87 ADC-channel/keV. %This denotes that the upper range of 
Temperature variations during above calibration rubs expressed as standard deviations ranged from 0.5 to 2.3$^{\circ}$C. 
Mean temperature values of each module are shown in Fig.~\ref{fig:pmtemp}. 
%/data2/polardata/calibration/Na22/bank20_201708/temperature2D_bank20_sept29.pdf

\begin{figure}[H]
\begin{center}
 \includegraphics[width=0.8\textwidth]{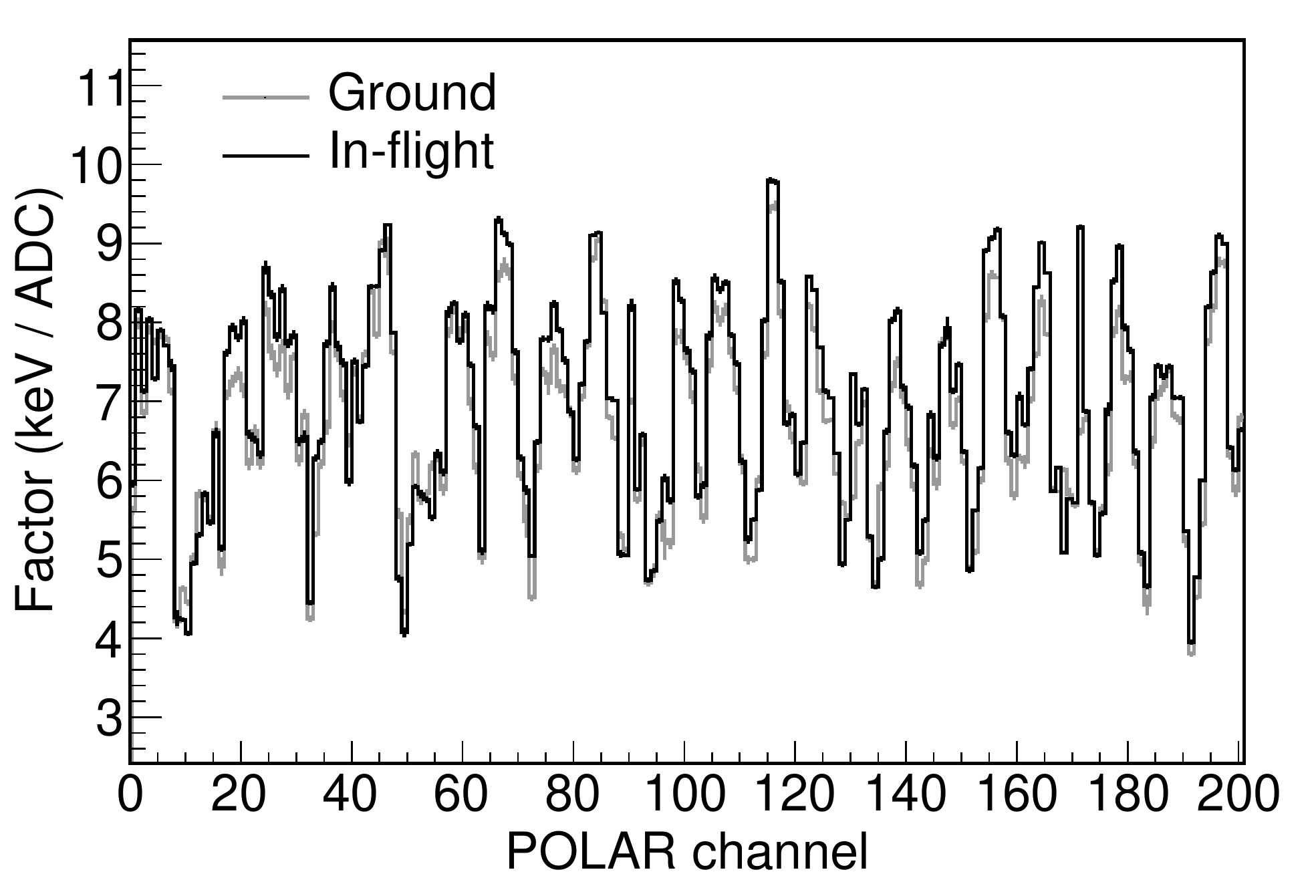}
\caption{
Comparison of the on-ground and in-flight energy conversion factors for the first 200 POLAR channels.
}
\label{fig:compare}
\end{center}
\end{figure}

\begin{figure}[H]
\begin{center}
 \includegraphics[width=0.8\textwidth]{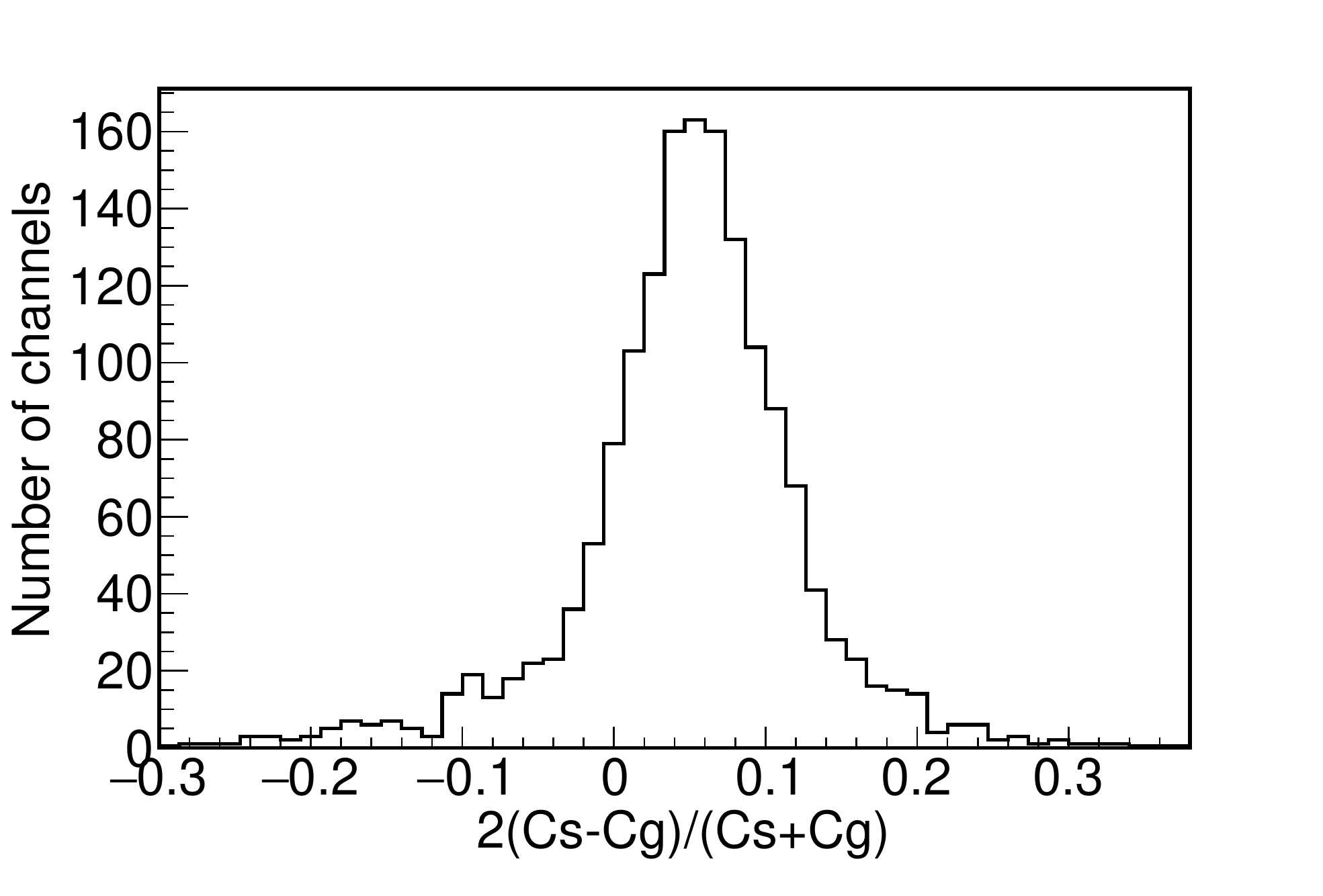}
\caption{
Distribution of relative differences between on-ground and in-flight energy conversion factors for all 1600 channels. 
The mean and the standard deviation values are equal to 4.9\% and 7.4\% respectively. Note that the presented results have no temperature corrections.
}
\label{fig:compare2}
\end{center}
\end{figure}

Energy conversion factors from the in-flight calibrations can be compared with the laboratory results for
the same HV and threshold settings.
For this purpose the data from calibration runs performed at the launch site in July 2016 was analysed. 
The laboratory and space data were processed using the same procedures and methods.
The mean temperature of POLAR modules during on-ground calibrations was just about 5$^\circ$C higher than in space.
Fig.~\ref{fig:compare} shows both
the laboratory  $c_{\rm g}$ and the in-flight energy conversion factors $c_{\rm s}$ of the
first 200 POLAR channels presented without temperature effect corrections.  
The distribution of the relative differences of the conversion factors, i.e. $2\cdot(c_{\rm s}-c_{\rm g})/(c_{\rm s}+c_{\rm g})$, for all 1600 channels is plotted in Fig.~\ref{fig:compare2}.  
The mean in-flight calibration factor is just about 4.9\% larger than the one measured on the ground factors. 
The mean difference is mostly due to temperature effects. 
Temperature variations during calibrations expressed as standard deviations range from 0.5 to 2.3$^{\circ}$C. 
After applying temperature corrections the individual differences are about 7.4\% (standard deviation).
They might be caused by vibrations during the launching.
The difference between values of the mean calibration factors almost vanish. 
More detailed studies of temperature effects are presented in the next section. 

%Based on above results it can be seen that the detector responses did not change significantly after the launch.

%Processing draw_ground_space_comparision.cc...
%root [1] Error in <TClingCallFunc::IFacePtr(kind)>: Attempt to get interface while invalid.
%Error in <TClingCallFunc::Exec(address, interpVal)>: Called with no wrapper, not implemented!
%Info in <TCanvas::Print>: pdf file /home/xiaohl/svn_psi_polar/software/energyResolutionFit/draw_200ch_compare.pdf has been created

%/home/xiaohl/svn_psi_polar/software/energyResolutionFit/spaceGroundCompare.pdf 

\subsection{Temperature effects}
\begin{figure}[H]
\begin{center}
 \includegraphics[width=0.7\textwidth]{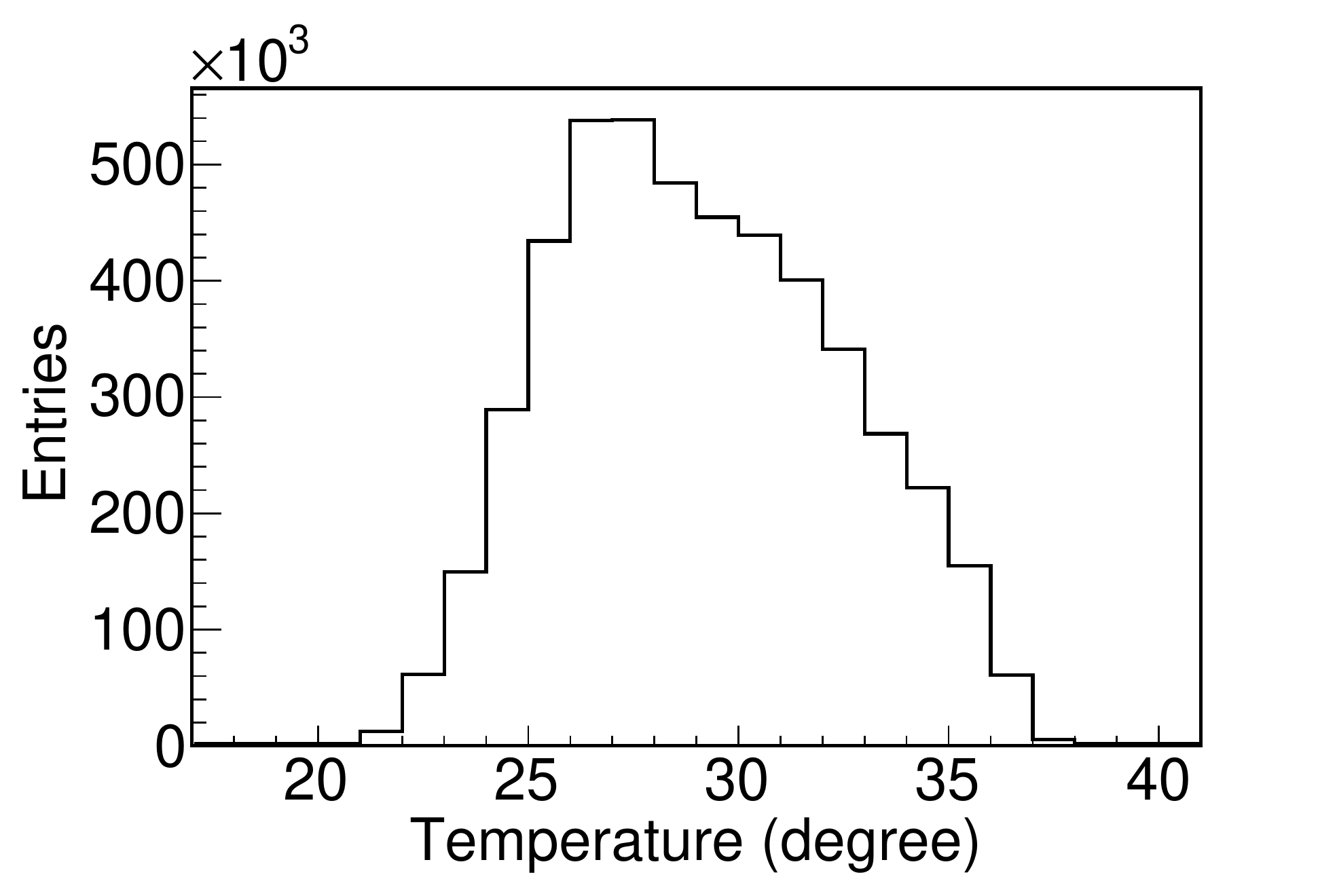}
\caption{Distribution of temperature of module 25 for the first six months  in space.  
Temperature values are readout every second. 
}
\label{fig:temperature}
\end{center}
\end{figure}

\begin{figure}[H]
\begin{center}
 \includegraphics[width=0.45\textwidth]{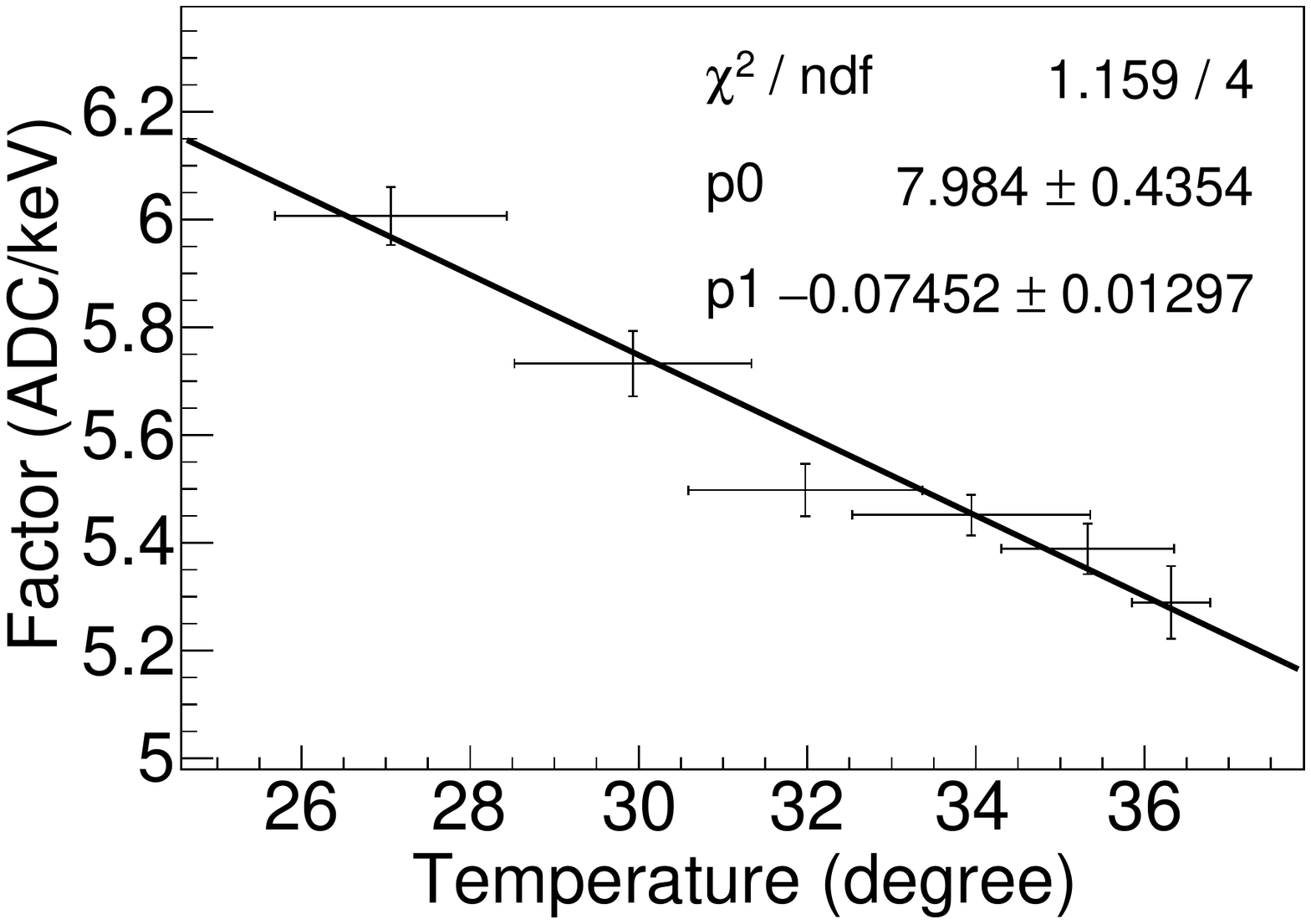}
 \includegraphics[width=0.45\textwidth]{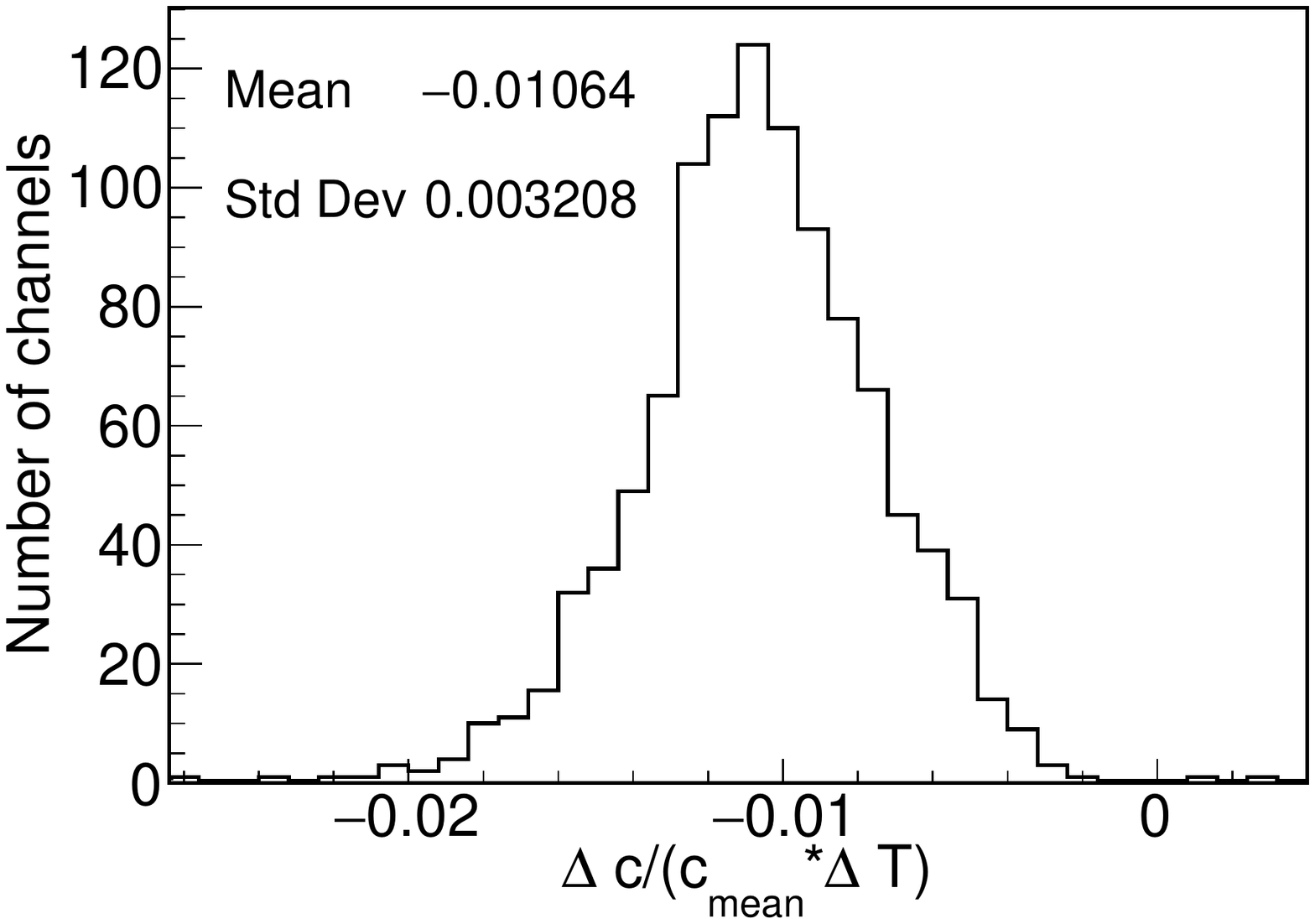}
\caption{Energy conversion factors as a function of temperature for the POLAR channel 12 (left) and distribution of temperature coefficients (right). 
}
\label{fig:tempfit}
\end{center}
\end{figure}

%As can be seen in the left panel of Fig.~\ref{fig:temperature}, 
The module temperature varies a 
few degrees along the spacecraft orbit largely due to differences in its illumination by the sun.
To study temperature dependence of energy conversion factors, we again used the same calibration data as before applying it for each POLAR channel. 
All coincidence events in the level-2B datasets were divided into six temperature groups.
For each group the corresponding energy spectra were prepared and used to determine their mean energy conversion factors. 
The left panel of Fig.~\ref{fig:tempfit} shows an example of the energy conversion factors for each temperature group. 
Both the mean temperature and its standard deviation for each group were calculated and depicted in the Figure. 
The same temperature values were assumed for all channels in the module. 
The energy vs. temperature coefficient has, as expected a negative temperature dependence \cite{hamamatsu}.
Its value was fitted for each channel using a linear function. 
The relative temperature coefficient $\alpha_{\rm T}$, i.e. the relative change of energy 
conversion factor per degree, was obtained with the formula $\alpha_{\rm T}= p_1/\bar{c}$, 
where $p_1$ is the slope from the linear fit and $\bar{c}$ is the mean value of the energy conversion factor. 
The distribution of temperature coefficients for about 1100 channels with sufficient amount of data 
is shown in the right panel of Fig.~\ref{fig:tempfit}. 
The mean coefficient from the fit is -1.06\% per degree. 
For the first six months of the POLAR operation in space a typical temperature variation
for a single module was equal to about 10$^\circ$C as shown in Fig.~\ref{fig:temperature}. 
Thus, typical drifts of the gain value due to temperature effects are of about 10\% only.

\subsection{Energy calibration with the high HV settings}

For higher values of HV it is impossible to conduct direct calibrations for channels with Compton edge positions outside of the ADC range.
Therefore, the energy  calibration for these channels relies on the indirect method described in Refs.~\cite{hlieee} and ~\cite{zhangxfgain}. 
The principle of the method is briefly introduced as below.

The MAPMTs of POLAR operate using an equally distributed voltage divider. 
Their gain factor in function of the applied HV value ($V$) is given by the following equation \cite{hamamatsu}:
\begin{equation}
 G= a^n(\frac{V}{n+1})^{kn},
 \label{eq:gain}
\end{equation}
where $a$ is a constant, $n$ is the number of dynode stages and
$k$ is a constant determined by the structure and material of the PMT. 
The typical value of $k$ is equal to 0.7 \cite{hamamatsu}. 
In the case of POLAR, $n$ is equal to 12.  
According to Eq. (\ref{eq:gain}), the gain $G$ is proportional to the $kn$-th power of the HV value.
It is reasonable to assume that the energy response of POLAR detector close to its typical operating 
condition is linear. 
Therefore the energy conversion $c$ (in units of ADC channel / keV) can be given by 
\begin{equation}
 c(V)=\frac{ E_{\rm meas}}{E_{\rm vis}}=b G=\alpha (\frac{V}{n+1})^{kn},
 \label{eq:cal}
\end{equation}
where $b$ is a constant, $V$ is the HV value, $\alpha$ is equal to $b a^n$, 
$E_{\rm vis}$ is the energy deposition (in units of keV) and $E_{\rm meas}$ is 
the recorded energy deposition (in units of ADC channel). 
Again, for POLAR PMT, $n$ is equal to 12.  
Eq.~(\ref{eq:cal}) can be parametrized using calibration data taken with several HV settings. 
Thus, the energy conversion factors for settings with higher values of HV can be determined using 
the parametrized function in the extrapolated range.

\begin{figure}[H]
\begin{center}
\includegraphics[width=0.47\textwidth]{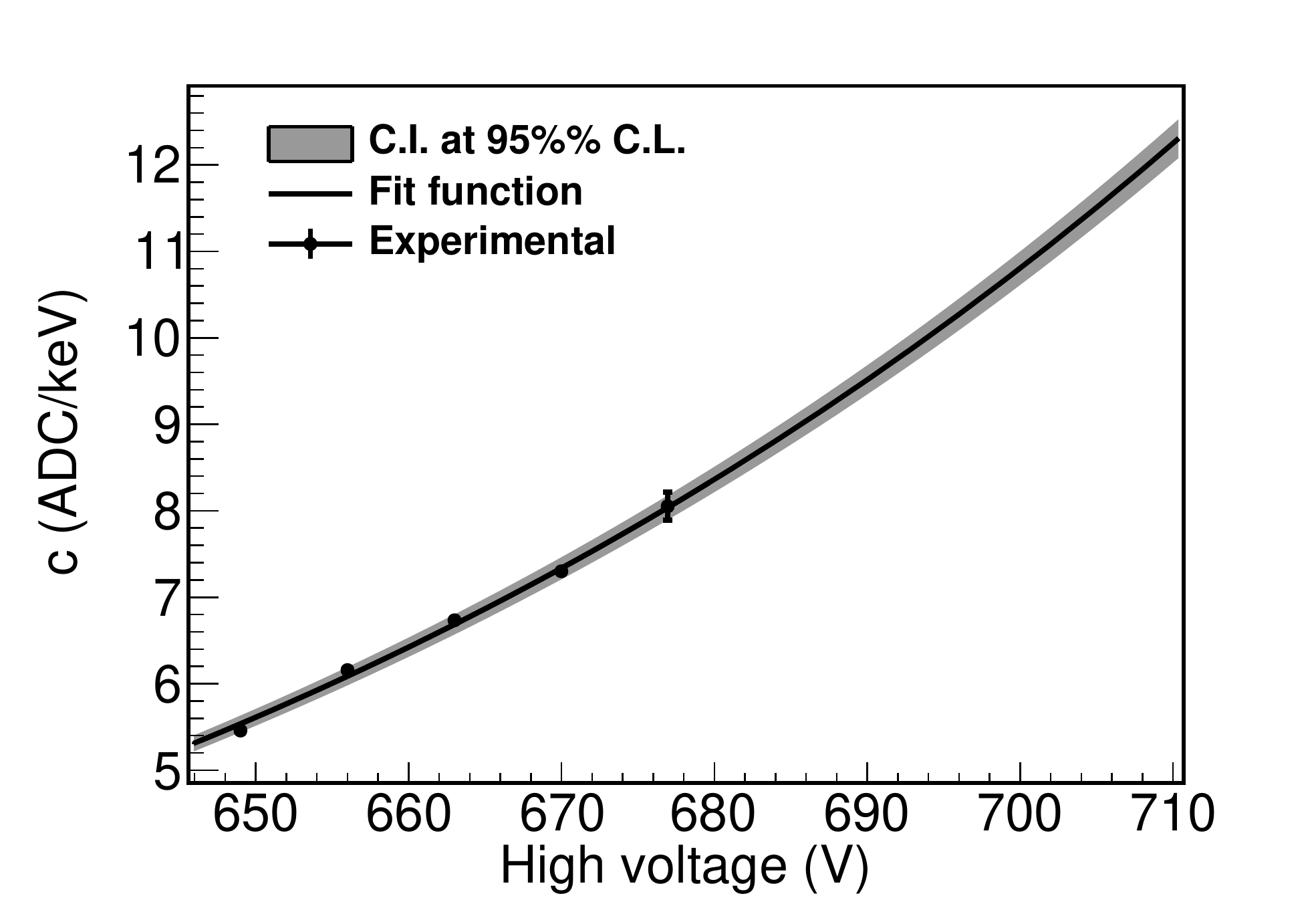} 
\includegraphics[width=0.47\textwidth]{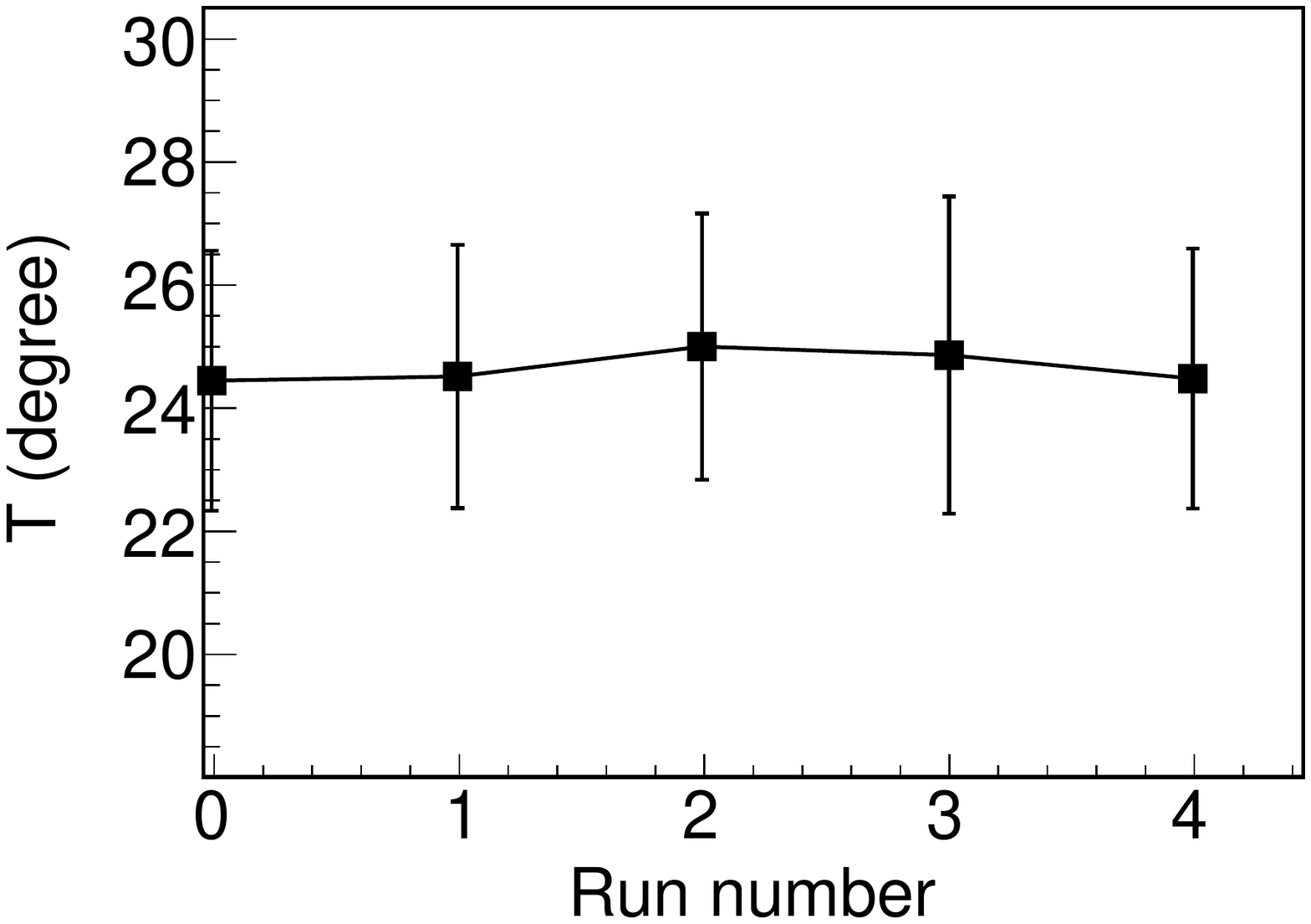} 
\caption{Left: Energy conversion factors  vs. HV values presented for the POLAR channel 4. 
A data fit with Eq.~(\ref{eq:cal}) together with the confidence intervals (at 95\% C.L.) are also shown.
Fit parameters $k$ and $\alpha$ are equal to 0.737 and $5.329\times 10^{-15}$ respectively. 
Right: Mean temperatures of POLAR module No. 1 during calibration runs.
}
\label{fig:inflight}
\end{center}
\end{figure}
%/data2/polardata/calibration/2018_calibration_using_unfolding/hv_vs_gain
\begin{figure}[H]
\begin{center}
\includegraphics[width=0.7\textwidth]{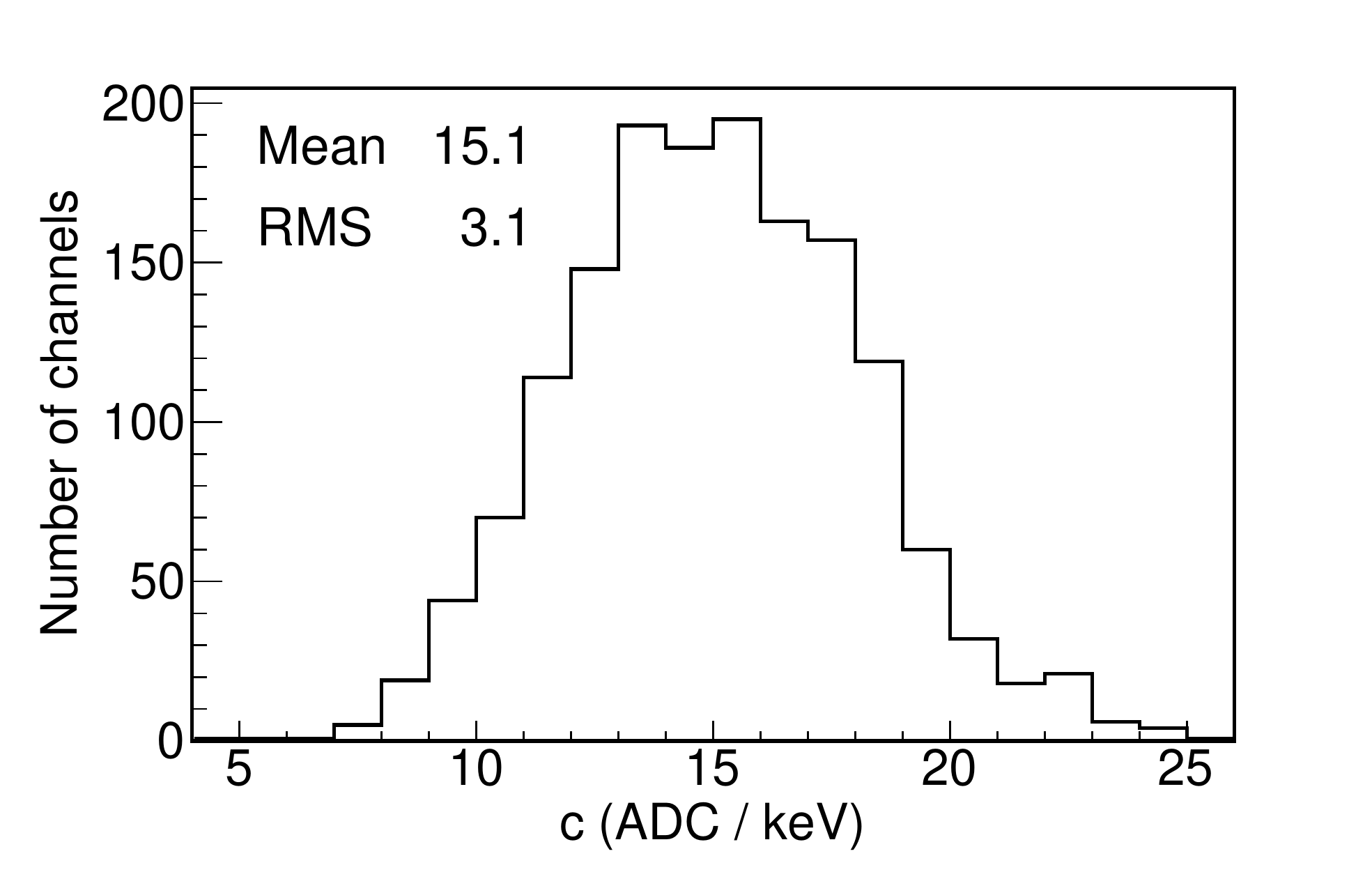} 
\caption{Distribution of the energy conversion factors calculated using Eq.~(\ref{eq:cal}) and parametrized for the high values of HV setting.
The mean energy conversion factor is equal to 15.1 ADC channel / keV.
}
\label{fig:bank50}
\end{center}
\end{figure}
%/data2/polardata/calibration/2018_calibration_using_unfolding/hv_vs_gain/bank50_factor.pdf
For the purpose of extended calibrations one prepared five HV settings. 
They were used to study dependence of the energy conversion factor on the HV value.
The generated HV values differed by -21 V, -14 V, -7 V, 0V and +6 V relative to the basic calibration settings. 
Several in-flight calibration runs have been  performed using above settings to date. 
The left panel of Fig.~\ref{fig:inflight} shows an example of the  energy conversion factors as a function of HV values with extended calibration data taken in November 2016.
The fit to the data done using Eq.~(\ref{eq:cal}) together with the confidence intervals (at 95\% C.L.) are also shown in the figure.  
Temperature corrections were not applied as the differences between the mean temperatures of the modules were negligible during all calibration runs. 
As an example, the mean temperature of module No. 1 is plotted in the right panel of Fig.~\ref{fig:inflight}. 

\begin{figure}[H]
\begin{center}
\includegraphics[width=0.7\textwidth]{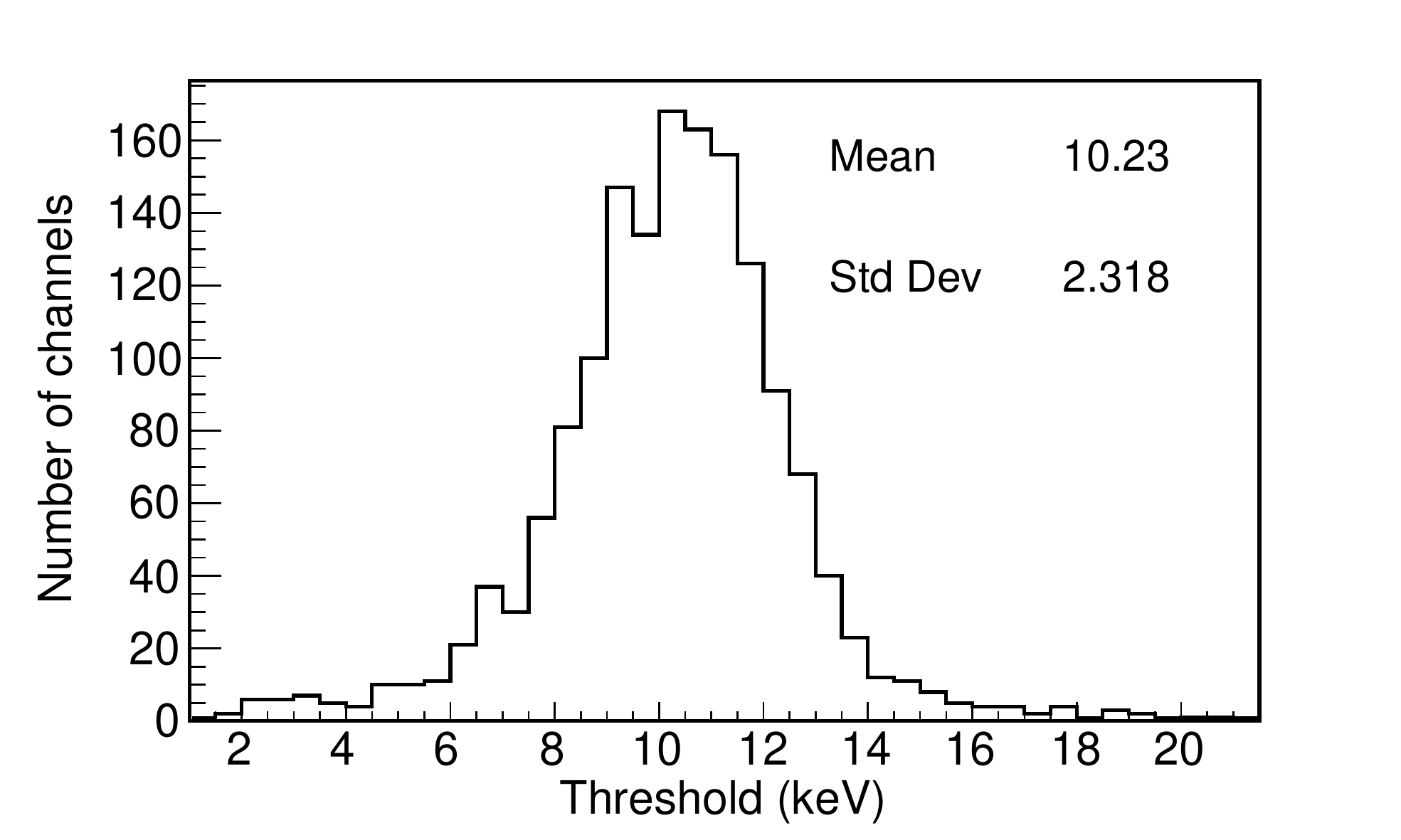} 
\caption{%An example of the threshold fit (left) and 
Distribution of threshold values for the high HV settings. The HV values are shown in Fig.~\ref{fig:na22hv}.
}
\label{fig:bank50threshold}
\end{center}
\end{figure}
%/data2/polardata/calibration/2018_calibration_using_unfolding/threshold/threshold_bank50_840.pdf has been created

Most of the time during its first six months in space POLAR operated with high values of its HV settings. 
They are shown in Fig.~\ref{fig:na22hv}.  
The energy conversion factors were obtained from the calibration data taken in November 2016. 
Their distribution for all 1600 POLAR channels is shown in Fig.~\ref{fig:bank50}.  
Note that the final calibration data for each detected GRB have  to be corrected for temperature effects and 
long-term drifts in the detector performance. 
Thus, any details of dedicated calibration routines applied for specific GRBs will be presented in future papers.

The mean threshold value with the basic HV setting was about 22 keV. 
The corresponding value of the discriminator threshold voltage (vthr) could not be set lower because of electronic noise in the readout system.   
Lower energy thresholds corresponding to higher values of HV are needed to accumulate more events especially from weaker GRBs and 
to improve the signal to background ratio for GRB detection.

In order to estimate values of the low energy thresholds, we constructed for each POLAR channel the energy spectra with all triggered hits.  
A sharp cut-off on the left side of the spectrum is related with the low energy threshold. 
Its value (in units of ADC channel) was chosen to be given by the position of the half-maximum of the cut-off. 
Fig.~\ref{fig:bank50threshold} shows the distribution of the threshold values for the high HV setting obtained using the energy conversion factors.  
Neither quenching effects nor crosstalk corrections was taken into account yet as final refinement of the calibration data for each detected 
GRB is still ongoing.

\section{Conclusion}
POLAR is a compact, wide-field of view, space-borne detector devoted for precise measurements of the linear polarization of hard X-rays.
The instrument is optimized for X-ray detection from GRBs and solar flares in the energy range from 50 keV to 500 keV.
POLAR was successfully launched on-board the Chinese space laboratory TG-2 on 15th September in 2016.
Energy calibrations of POLAR in space are performed using its four low activity $^{22}$Na calibration sources.
The method relies on the Compton edge measurements in the coincidence energy spectra of the collinear 511 keV photon pairs from positron annihilation.
The measurements are taken with pre-defined basic HV and threshold settings.
The method was extensively studied and optimized using Monte Carlo simulations and carefully verified in the laboratory before the launch.
During in-flight calibrations several HV and threshold settings were applied in regular time intervals.
Energy calibration factors were determined and compared with Monte Carlo simulations and on-ground data.
Results comparison between laboratory measurements and in-space calibration runs shows only a few percent deviations in the energy conversion factor values.
Additional correction factors for temperature effects were also determined and found to be on a level of one percent only.
Energy conversion factors applied during GRB measurements were determined using extrapolation procedures
based on the extended calibration runs. They were performed at different HV values applying also suitable temperature corrections.
Results from the in-flight calibration procedure shows high level of consistency with laboratory measurements.
Regular calibration runs also proved stability of POLAR instruments for the first 6 months of its operation in space.
Collected calibration database allows for precise determination and fine tuning of energy calibration factors for each GRB detected by POLAR.

\section*{Acknowledgements}
We gratefully acknowledge financial support from the National Basic Research Program (973 Program) of China (Grant No. 2014CB845800), 
the National Natural Science Foundation of China (Grant No. 11403028), Swiss National Science Foundation, Swiss Space office (ESA PRODEX program)
 and National Science Center Poland (Grant 
 No. 2015/17/N/ST9/03556).

%%%%%%%%%%%%%%%%%%%%%%%
\bibliographystyle{model1-num-names}
%\bibliography{calibration}

\bibliography{calibration}

\end{document}